\documentclass[conference]{IEEEtran}
\usepackage{fst}
\newcommand{\GL}{}

\newcommand{\EBY}{}
\newcommand{\BM}{}

\makeatletter
\newcommand{\thickhline}{%
    \noalign {\ifnum 0=`}\fi \hrule height 1pt
    \futurelet \reserved@a \@xhline
}
\makeatother

\begin{document}

\title{Protograph-Based LDPC Code Design for Ternary Message Passing \GL{Decoding}}

\author{\IEEEauthorblockN{Emna Ben Yacoub$^{\dagger,\ddagger}$, Fabian Steiner$^\ddagger$, Balazs Matuz$^\dagger$, Gianluigi Liva$^\dagger$}
\IEEEauthorblockA{$^\dagger$Institute of Communications and Navigation, German Aerospace Center (DLR), Germany\\$^\ddagger$Institute for Communications Engineering, Technical University of Munich, Germany\\Email: \{emna.ben-yacoub, fabian.steiner\}@tum.de, \{balazs.matuz, gianluigi.liva\}@dlr.de}}

\markboth{}{}%


\maketitle

\begin{abstract}
\EBY{A} ternary message passing (TMP) decoding algorithm for low-density parity-check  \GL{codes} \EBY{is developed}. \BM{All messages exchanged between}  variable and check nodes \EBY{have} a ternary alphabet, \EBY{and the} variable nodes exploit soft information from the channel. A density evolution analysis is developed for unstructured and protograph-based ensembles. For unstructured ensembles  the stability condition is  derived. \EBY{Optimized} ensembles for TMP decoding \BM{show asymptotic gains of up to \SI{0.6}{dB}} \EBY{with respect to} ensembles optimized for binary message passing decoding. Finite length simulations of \GL{codes from} TMP-optimized ensembles show gains of up to \SI{0.5}{dB} \GL{under} TMP compared to
protograph-based codes designed for unquantized \GL{belief propagation} decoding.


\end{abstract}


\section{Introduction}

With the advent of iterative and soft information based \ac{FEC} schemes \ac{LDPC} codes \cite{gallagerPhD}
found  widespread use in many modern communication standards\GL{, e.g., for} digital 
video broadcasting~\cite{etsi2009dvb}, optical communications~\cite{ieee802.3ca} 
and wireless local area networks~\cite{ieee802.11}. \EBY{Recently}, protograph\GL{-based \cite{thorpe_protograph}, rate-compatible} \ac{LDPC} 
codes~\cite{chen_protograph-based_2015} were chosen as the
\ac{FEC} solution for \EBY{the} enhanced mobile broadband \GL{(eMBB) \BM{service} of the 3GPP 5G standard \cite{richardson_design_2018}}.
Because of increasing data rate requirements, the need for low-complexity and
high throughput decoding algorithms is \EBY{acute}. For \ac{LDPC} codes,
the data flow between the \ac{VN} and \ac{CN}
component codes during one iteration of the \ac{BP} decoding algorithm is a major \BM{source of} complexity.
\GL{The data flow is linear in the number of quantization bits of the messages \cite{smith_staircase_2012}.}
For applications with very high throughput requirements such as optical
communications, the messages are usually quantized to four bits to alleviate this
problem~\cite{schmalen_spatially_2015}.
In his thesis, Gallager
presents algorithms A and B which operate with binary messages. In \cite{lechner_analysis_2012}, the authors develop an improved
algorithm with binary messages, \BM{referred to as \ac{BMP} decoding}, which allows to exploit the channel soft information.

\GL{In this paper, we} extend the work of \cite{lechner_analysis_2012} by allowing a ternary message alphabet. \GL{\BM{We introduce} an erasure as a third message value to denote complete
uncertainty about the  \BM{respective bit value}. The resulting algorithm is \BM{referred} to as \ac{TMP} decoding. \ac{TMP} decoding closely resembles algorithm~E from \cite{richardson_capacity}, \EBY{except that \ac{TMP} exploits} soft information available at the channel output.}

The motivation for \ac{TMP} \GL{decoding}  is twofold. First, previous works have shown
significant gains if erasures are allowed in the decoding process~\cite{richardson_capacity,forney_generalized_1966,kwon_optimal_2008}. Second, \ac{TMP} \GL{decoding \EBY{applies} to \ac{LDPC} codes with state \acp{VN} which \EBY{can improve} decoding thresholds. \BM{Examples are accumulator-based constructions from \cite{divsalar_capacity-approaching_2009} or recently standardized 5G codes \cite{richardson_design_2018}}.  In fact, one may argue that  \ac{TMP} messages are represented by $2$ bits, and hence one quantization level is lost by restricting the message alphabet to be ternary. However, whenever state \acp{VN} are used, their initial \ac{LLR} \EBY{is} set to zero. \BM{Although} this issue might be mitigated by introducing a non-trivial message passing schedule, \ac{TMP} \EBY{is a} simple \EBY{method} to account for the lack of channel observations at the input of state \acp{VN}. \EBY{This} argument \BM{holds also for} rate compatible \ac{LDPC} code constructions relying on puncturing of low-rate mother codes \cite{zhang2007structured}.}

In this paper, we describe \GL{the \ac{TMP} decoding algorithm} for both unstructured and protograph \GL{\ac{LDPC}} ensembles. \GL{We \EBY{then} develop the exact \ac{DE} analysis} to compute decoding thresholds\BM{.  
We} develop the stability condition for unstructured ensembles
and discuss its difference to \cite{lechner_analysis_2012}. Further, we compare the decoding thresholds \GL{under}
\ac{BMP} and \ac{TMP} \GL{decoding} for optimized protograph  ensembles \GL{targeting} \EBY{code} rates from $2/3$  to $9/10$.

The paper is organized as follows. In Sec.~\ref{sec:prelimin} we briefly discuss the system model and introduce \EBY{notation}. Sec.~\ref{sec:tmp} \BM{presents} the \ac{TMP} decoding algorithm. In Sec.~\ref{sec:de}, we introduce the \ac{DE} analysis of \ac{TMP} decoding for unstructured and protograph-based \ac{LDPC} ensembles. Numerical results are discussed in Sec.~\ref{sec:results}. \GL{Conclusions follow in Sec.~\ref{sec:conclusions}.}

\section{Preliminaries}\label{sec:prelimin}

\subsection{System Model}
\label{sec:biawgnc}
We consider the \ac{biAWGN} channel with input alphabet $ \cX  = \{-1, +1 \}$. The channel output is $ Y = X + N $, where $ N $ is  Gaussian random noise with zero mean and variance $\sigma^{2} $. The \BM{channel quality} is defined in terms of $ E_{\tb}/N_{0} $, with $ E_{\tb} $ being the energy per information bit and $ N_{0} $ the single-sided noise power spectral density.

\subsection{Extrinsic Channel}
\label{sec:beec}

\BM{The messages passed in an iterative decoder can be modelled \EBY{as the output of} an extrinsic channel to which the respective \acp{VN} and \acp{CN} are connected \EBY{\cite[Fig.~3]{ashikhmin_EXIT}}. For a \ac{TMP} decoder the extrinsic channel is a} \ac{BEEC} with input alphabet $ \cX  = \{-1, +1 \}$, output alphabet $ \cZ  = \{-1, 0, +1 \}$, where $ 0$ corresponds to an erasure.
Let $ \theta $ and $ \epsilon $ be the \EBY{respective} error and erasure probabilities of this channel. The channel \ac{LLR} of the \ac{BEEC} is
\BM{
\begin{align}\label{eq:LLRBEEC}
   L(z) &=  \ln \left[ \dfrac{\Pr\left\{ Z = z|X= +1\right\} }{\Pr\left\{Z=  z|X= -1\right\}}\right] \\
   &= \underbrace{\ln \left( \dfrac{ 1 - \theta - \epsilon }{\theta} \right)}_{D} \cdot z \label{eq:2:rel}
\end{align}
where $ D $ denotes the message reliability.}

\subsection{Low-Density Parity-Check Codes }
\label{sec:ldpc}

Binary \ac{LDPC} codes are binary linear block codes defined by an $ m \times n $ sparse parity-check matrix \GL{$\bm{H}$}. The code dimension is $\GL{k} \leqslant n - m$. The Tanner graph of an \ac{LDPC} code is a bipartite graph $G = (\cV \cup \cC, \cE)$ consisting of $n$ \acp{VN} and $m$ \acp{CN}. The set $\cE$ of edges contains the element\GL{s
$e_{ij}$, where $e_{ij}$ is an edge between \ac{VN} $V_j\in\cV$ and \ac{CN} $C_i\in\cC$. Note that $e_{ij}$ belongs to the set $\cE$ if and only if the \BM{parity-check} matrix element \EBY{$h_{ij}$} is equal to $1$.}
The sets $\cN(V_j)$ and $\cN(C_i)$ denote the neighbors of \ac{VN} $V_j$ and \ac{CN} $C_i$,
respectively. The degree  \GL{of a \ac{VN} $V_j$ is denoted by $d_{\tv,j}$ and it is the cardinality of the set $\cN(V_j)$. Similarly, the degree of a \ac{CN} $C_i$ is denoted by $d_{\tc,i}$ and it is the cardinality of the set $\cN(C_i)$.}

\medskip

\subsubsection{Unstructured Ensembles} The \ac{VN} edge-oriented degree distribution \GL{polynomial} of an \ac{LDPC} code graph is given by 
$
    \lambda(x)= \sum_{j}\lambda_{j}x^{j-1}
$
where $ \lambda_{j} $ corresponds to the fraction of edges incident to \ac{VN}s with degree \BM{$ j $}.
Similarly, the \ac{CN} edge-oriented degree distribution polynomial is given by
$
\rho \left(x\right) = \sum_{i}\rho_{i}x^{i-1}
$
where $ \rho_{i} $ corresponds to the fraction of edges incident to \ac{CN}s with degree $ i $.
An unstructured irregular \ac{LDPC} code ensemble $ \mathcal{C}_{n}^{\lambda,\rho} $ is the set of all \ac{LDPC} codes with block length $ n $ and degree distributions $ \lambda\left( x\right) $ and $ \rho\left( x\right) $.

\medskip

\subsubsection{\GL{Protograph Ensembles}} For practical purposes it is \GL{often} worthwhile to impose more
structure on a given \GL{\ac{LDPC} code ensemble. Examples of structured \ac{LDPC} code ensembles} are \ac{MET}~\cite{richardson_MET} \BM{and} protograph-based ensembles~\cite{thorpe_protograph}.
\GL{Protograph-based ensembles} are defined via a \GL{(typically small)} base matrix $\vB$ of dimension \GL{$m_0 \times n_0$} and elements in $\{0,1,\ldots,S\}$.
A base matrix may also be represented as bipartite graph $ P $ (called protograph) as described before \GL{for the parity-check matrix case. However, since the elements of the base matrix are not strictly binary, parallel edges (in number corresponding to the multiplicity of the corresponding base matrix element) are allowed.} The Tanner graph of an \ac{LDPC} code is obtained via \BM{lifting: through copy-and-permute operations a} number of copies of the protograph \BM{is generated and} their edges are permuted such that connectivity constraints imposed by the base matrix are maintained \cite{thorpe_protograph}. A protograph-based \ac{LDPC} \BM{code} ensemble $ \cC_{n}^{P} $ is defined by the set of \BM{length-$n$} \ac{LDPC} codes whose Tanner graph is obtained by lifting $P$.

\section{Ternary Message Passing Algorithm}
\label{sec:tmp}

We denote by  $ m_{C \to V}^{( \ell) }  $  the message sent from \ac{CN} $ C $ to its neighboring \ac{VN} $ V $. Similarly, $ m_{V \to C}^{(\ell) }  $ is the message sent from \ac{VN} $ V $ to \ac{CN} $ C $ at the $\ell$-th iteration. The alphabet of the exchanged messages between the \BM{\acp{CN} and \acp{VN}} is $ \cM=\{ -1, 0, +1 \} $, where $ 0 $ corresponds to an erasure. 

\BM{Initially}, each \ac{VN} computes the \ac{LLR} \[\GL{L_{\tch} = \frac{2}{\sigma^2}y}\]
of its channel observation and passes a \BM{quantized value} to its neighboring \acp{CN}. Hence, for all $ C \in \cN( V) $ \EBY{we have}
\begin{equation}\label{eq:VNupdatel0}
m_{V \to  C}^{( 0) } = f( L_{\tch}) 
\end{equation}
where the quantization function $ f : \mathbb{R} \to \cM $ converts soft \BM{$L$-values} to ternary messages and is defined as
\begin{align} \label{eq:functionf}
f (x) =
\begin{cases}
+1 & x > a \\
0 & -a \leqslant x\leqslant a \\
-1 & x < -a.
\end{cases}
\end{align} 
\BM{\EBY{We} choose $a$ \EBY{to minimize the decoding threshold.}}

At the $\ell$-th iteration, \ac{CN} $ C $ sends to its neighboring \ac{VN} $ V $ the product of the messages that it received from the other neighboring \ac{VN}s, i.e., \EBY{we have}
\begin{equation}\label{eq:CNupdate}
m_{C\to V}^{(\ell) }  = \prod\limits_{V' \in \cN (C) \setminus V} m_{V'\to C}^{({\ell-1}) }.
\end{equation}

Each \ac{VN} converts the channel output and the incoming \ac{CN} messages to $L$-values and passes the quantization of the \EBY{$L$-values} to its neighboring \acp{CN}. We get
\GL{\begin{equation}\label{eq:VNupdateAlgoS1}
m_{V\to C}^{(\ell) }  = f\left( L_{\tch} + {L_{\tin}^{(\ell)}}\right)
\end{equation}
with
\begin{equation} \label{eq:3:VN}
L_{\tin}^{(\ell)}:=\sum\limits_{C' \in \cN\left( V\right) \setminus C }D^{(\ell) }_{C'V} m_{C'\to V}^{(\ell) }.
\end{equation}}
\EBY{For the estimation of its modulated codeword bit} , each \ac{VN} computes 
\begin{equation}\label{eq:APPalgoS}
\EBY{\hat{x}_{V}^{(\ell) }}  = \sign \left( L_{\tch} + \tilde{L}_{\tin}^{(\ell)} \right)
\end{equation}
with
\begin{equation} \label{eq:3:VN_app}
\tilde{L}_{\tin}^{(\ell)}:=\sum\limits_{C' \in \cN\left( V\right)}D^{(\ell) }_{C'V} m_{C'\to V}^{(\ell) }.
\end{equation}
\BM{In~\eqref{eq:3:VN} and \eqref{eq:3:VN_app}, $D^{(\ell) }_{C'V}$ is a weighting factor whose value is determined as part of the decoder design. Starting from~\eqref{eq:2:rel} the reliability of a message from  $ C $ to $ V $ on the  \ac{BEEC}  is}
\begin{equation}\label{eq:DCV}
D_{CV}^{(\ell)} = \ln\left( \dfrac{1-q_{0}^{(\ell) }( C,V) -q_{-1}^{(\ell) }(C,V) }{q_{-1}^{(\ell) }(C,V)}\right).
\end{equation}
For unstructured ensembles, $ q_{0}^{(\ell) }( C,V) $ and $ q_{-1}^{(\ell) }( C,V)$ are the average of the erasure and error probabilities over the \ac{CN} edge-oriented degree distribution, respectively. For protograph-\ac{LDPC} ensembles, $ q_{0}^{(\ell) }( C,V) $ and $ q_{-1}^{(\ell) }( C,V)$ are respectively the erasure and  error probabilities of the message sent over an edge of the type defined by the pair $(C,V)$.

Note that $ q_{0}^{(\ell) }( C,V) $ and $ q_{-1}^{(\ell) }( C,V)$ can be estimated via \ac{DE} \BM{discussed in} Sec.~\ref{sec:de}. \GL{Note that the different $  D_{CV}^{(\ell)} $\EBY{,} as well as the quantization parameter $a$ of \eqref{eq:functionf}\EBY{,} are \BM{determined}  during the decoder design phase. Hence, their calculation does not contribute to the decoding complexity}.

\section{Density Evolution Analysis}\label{sec:de}
\label{sec:pexit}

\GL{We provide a \ac{DE} analysis for both unstructured and protograph-based \ac{LDPC} code ensembles. In the following, $L_{\tmax}$ denotes the maximum number of iterations used in the derivation of the decoding thresholds.}

\subsection{Unstructured \ac{LDPC} Code Ensembles}

\EBY{Let} $ p_{0}^{(\ell)} $ and $ p_{-1}^{(\ell)} $ \EBY{be} the erasure and error probabilities of \ac{VN} messages at the $\ell$-th iteration. Similarly, $ q_{0}^{(\ell)} $ and $ q_{-1}^{(\ell)} $ are the erasure and error probabilities of \ac{CN} messages. \GL{In the limit of $n\rightarrow \infty$,} \ac{DE} can be summarized \BM{as follows}.

\begin{enumerate}
	\item \textbf{Initialization.}
	\GL{\BM{Conditioned} on $ X = +1$ (all-zero codeword)}, the \BM{channel \acp{LLR}} are Gaussian \acp{RV} with mean  $ 	\mu_{\tch}= 4RE_{\tb}/N_{0} $ and variance $ \sigma_{\tch}^{2}= 2 \mu_{\tch} $. Therefore, \GL{recalling \eqref{eq:VNupdatel0}}, \EBY{we have}
	
	
	

	\begin{align} 
	\begin{split}
	p^{(0)}_{0}&= \Pr\left\{ -a \leqslant L_{\tch} \leqslant a \right\}  \\
	&= Q\left( \dfrac{-a+\mu_{\tch}}{\sigma_{\tch}}\right)  -  Q\left( \dfrac{a+\mu_{\tch}}{\sigma_{\tch}}\right) 
	\end{split}\\
	p^{(0)}_{-1}&= \Pr\left\{L_{\tch} < -a \right\} =  Q\left( \dfrac{a+\mu_{\tch}}{\sigma_{\tch}}\right)
	\end{align}
	
	where  the $Q$\BM{-}function is defined as

\begin{equation}\label{eq:Qfunction}
Q (x)  = \dfrac{1}{\sqrt{2\pi}} \int_{x}^{\infty} e^{-\frac{z^{2}}{2}} \dif{z}.
\end{equation}

	\item \textbf{For} \BM{$ \ell =1, 2, \ldots, L_{\tmax} $}\\[2mm]
     \textbf{Check to variable update}
	\BM{	\begin{align} 
		q^{(\ell)}_{0}  =& 1-\rho\left(1-p^{(\ell-1)}_{0}\right)\\ 
			q^{(\ell)}_{-1}  =& \frac{1}{2} \left[\rho\left(1-p^{(\ell-1)}_{0}\right)	-\rho\left(1 - 2p^{(\ell-1)}_{-1}-p^{(\ell-1)}_{0}\right)\right].
		\end{align}}
	\textbf{Variable to check update}
		\begin{equation}
         \begin{aligned}
          p_{0}^{(\ell)} =& \sum\limits_{d} \lambda_{d} \sum\limits_{m_{\tin}}\Pr\left\{M_{\tin}^{(\ell)} = m_{\tin}\right\} \times\\
          &\left[Q\left(\frac{-a+D^{(\ell)}m_{\tin}+\mu_{\tch}}{\sigma_{\tch}}\right) \right.\\
          &\left.-Q\left(\frac{a+D^{(\ell)}m_{\tin}+\mu_{\tch}}{\sigma_{\tch}}\right) \right]
         \end{aligned}
         \end{equation}
\begin{equation}
\begin{aligned}
p_{-1}^{(\ell)} =& \sum\limits_{d} \lambda_{d} \sum\limits_{m_{\tin}} \Pr\left\{M_{\tin}^{(\ell)} = m_{\tin}\right\}\times \\
&Q\left(\frac{a+D^{(\ell)}m_{\tin}+\mu_{\tch}}{\sigma_{\tch}}\right)
\end{aligned}
\end{equation}

where 
\begin{equation}\label{eq:D}
D^{(\ell)} := \ln\left( \dfrac{1-q_{0}^{(\ell) } -q_{-1}^{(\ell) }}{q_{-1}^{(\ell) }}\right)
\end{equation}

and $ M_{\tin}^{(\ell) }$ is a \ac{RV} representing the sum of the $d-1$ incoming \ac{CN} messages at the $\ell$-th iteration. Moreover, \EBY{we have}
\begin{equation}\label{eq:PrMin}
\begin{aligned}
\Pr\left\{M_{\tin}^{(\ell)} = m_{\tin}\right\} =& \sum\limits_{\substack{u,v \\ u-v=m_{\tin}}}\binom{d-1}{u,v,d-1-u-v}\times \\
&\left(q_{-1}^{(\ell)}\right)^{v}  \left( q_{0}^{(\ell)}\right)^{d-1-u-v}\times \\
& \left(1- q_{-1}^{(\ell)}-q_{0}^{(\ell)}\right)^{u}.
\end{aligned}
\end{equation}


\end{enumerate}

The ensemble iterative decoding threshold $ (E_{\tb}/N_{0})^\star $ is defined as the minimum $ E_{\tb}/N_{0} $ for which $ p_{-1}^{(\ell)} \to 0 $ and $ p_{0}^{(\ell)} \to 0 $ as \GL{$ \BM{\ell} \to \infty$}.

\subsection{Protograph-Based LDPC Code Ensembles}

\EBY{Let} $ p_{0}^{(\ell)}( i,j) $  and $ p_{-1}^{(\ell)}( i,j)  $ \EBY{be} the erasure and error probabilities of the message sent from \GL{a \ac{VN} of type  $ V_{j} $} to \GL{a \ac{CN} of type $ C_{i} $}  at the $\ell$-th iteration on one of the $ b_{ij} $ edges connecting $ V_{j} $ to $ C_{i} $. Similarly, $ q_{0}^{(\ell)}( i,j) $ and $ q_{-1}^{(\ell)}( i,j) $ denote the erasure and error probabilities of the message sent from $ C_{i} $ to  $ V_{j} $ on one of the $ b_{ij} $ edges connecting   $ C_{i} $ to  $ V_{j} $. The error probability of the estimation at  the $\ell$-th iteration is denoted by $ P_{\tapp}^{(\ell)}(j) $.  \GL{In the limit of $n\rightarrow \infty$,} the protograph-based \ac{DE} analysis \cite{liva_protograph_2007} can be summarized in the following steps.
\begin{enumerate}
	\item \textbf{Initialization.}
	\BM{For $ j= 1, 2, \ldots, n_{0} $ and $ i=1, 2, \ldots, m_{0}$} with $ b_{ij} \neq 0 $, if $ V_{j} $ is \GL{a state \ac{VN}} 
	\begin{align}
	p^{(0)}_{0}(i,j) = 1 & &\text{and} & & p^{(0)}_{-1}(i,j) = 0.
	\end{align}
	
    	Otherwise,
	\begin{align} 
	\begin{split}
	p^{(0)}_{0}(i,j)=Q\left( \dfrac{-a+\mu_{\tch}}{\sigma_{\tch}}\right)  -  Q\left( \dfrac{a+\mu_{\tch}}{\sigma_{\tch}}\right) 
	\end{split}\\
	\begin{split}
	p^{(0)}_{-1}(i,j)=Q\left( \dfrac{a+\mu_{\tch}}{\sigma_{\tch}}\right).
	\end{split}
	\end{align}

	\item \textbf{For} $ \ell =1, 2, \ldots, L_{\tmax} $

\medskip
	
		\textbf{Check to variable update}
		
		\BM{For $ j= 1, 2, \ldots, n_{0} $ and $ i=1, 2, \ldots, m_{0}$},
		
		if $ b_{ij} \neq 0 $
		\begin{equation} \label{eq:q0AlgoSproto}
		q^{(\ell)}_{0}(i,j)  = 1 - \prod\limits_{b_{i,s}\neq 0}\left(1 - p^{(\ell-1)}_{0}(i,s) \right) ^{b_{i,s}-\delta_{sj}}
		\end{equation}
		\begin{equation} \label{eq:q_1AlgoSproto}
		\begin{aligned}
		&q^{(\ell)}_{-1}(i,j)  = \frac{1}{2} \left[\prod\limits_{b_{i,s}\neq 0}\left(1 - p^{(\ell-1)}_{0}(i,s) \right) ^{b_{i,s}- \delta_{sj}} \right. \\
		&-  \left. \prod\limits_{b_{i,s}\neq 0}\left(1 - 2p^{(\ell-1)}_{-1}(i,s)-p^{(\ell-1)}_{0}(i,s) \right) ^{b_{i,s}- \delta_{sj}}\right]
		\end{aligned}
		\end{equation}
		
		where $ \delta_{ij} $ is the Kronecker delta function.  
        
\medskip

		\textbf{Variable to check update}
		
		For \BM{$ j= 1, 2, \ldots, n_{0} $ and $ i=1, 2, \ldots, m_{0}$} with $ b_{ij} \neq 0 $, if $ V_{j} $ is punctured \EBY{we have} 
			\begin{align} 
				p^{(\ell)}_{0}(i,j) =& \Pr\left\{  -a \leqslant  L^{(\ell)}_{\tin} \leqslant a \right\} \\
				p^{(\ell)}_{-1}(i,j)=& \Pr\left\{   L^{(\ell)}_{\tin} < -a \right\}. 
					\end{align}
			Otherwise, \EBY{we have} 	
		\begin{equation} 
		\begin{aligned}
		p^{(\ell)}_{0}(i,j)=& \Pr\left\{  -a \leqslant L_{\tch}  + L^{(\ell)}_{\tin} \leqslant a \right\}  \\
		=& \sum\limits_{z} \Pr\left\{  L^{(\ell)}_{\tin} = z \right\} \left[Q\left( \dfrac{-a+z+\mu_{\tch}}{\sigma_{\tch}}\right) \right. \\
		&\left. - Q\left( \dfrac{a+z+\mu_{\tch}}{\sigma_{\tch}}\right) \right] \\
		p^{(\ell)}_{-1}(i,j)=& \Pr\left\{   L_{\tch}  + L^{(\ell)}_{\tin} < -a \right\}  \\
		=& \sum\limits_{z} \Pr\left\{  L^{(\ell)}_{\tin} = z \right\} Q\left( \dfrac{a+z+\mu_{\tch}}{\sigma_{\tch}}\right) 
		\end{aligned}
		\end{equation}
		
		where $ L^{(\ell)}_{\tin} $ is a \ac{RV} representing the sum of the \acp{LLR} of the $d_{\tv,j} - 1 $ \ac{CN} messages at the input of $ V_{j} $ at the $ \ell$-th iteration. We have

		\begin{equation}\label{eq:LinAlgS}
		\begin{aligned}
		\Pr\left\{  L^{(\ell)}_{\tin}= z \right\} &= \sum\limits_{\mathbf{u}, \mathbf{v}} \prod\limits_{b_{s,j}\neq 0} 
		{\textstyle \binom{b_{s,j} - \delta_{si}}{u_{s},v_{s},b_{s,j} - \delta_{si} - u_{s} - v_{s}}} \times \\
		&\left( 1 - 	q^{(\ell)}_{0}(s,j)  -	q^{(\ell)}_{-1}(s,j) \right) ^{u_{s}} \times \\
		& q^{(\ell)}_{0}(s,j)^{b_{s,j} -\delta_{si} - u_{s} - v_{s}} q^{(\ell)}_{-1}(s,j)^{v_{s}} 
		\end{aligned}
		\end{equation}
		\GL{where the outer sum is over all integer vector pairs $\mathbf{u},\mathbf{v}$  for which 
		\[\sum\limits_{e=1}^{m_{0}}w^{(\ell)}_{e,j}(u_{e}-v_{e}) = z
		\]
		with
		\begin{equation}\label{eq:wkjAlgoS}
		w^{(\ell)}_{e,j} := \ln\left( \dfrac{1-	q^{(\ell)}_{0}(e,j)  -	q^{(\ell)}_{-1}(e,j)}{	q^{(\ell)}_{-1}(e,j)}\right).
		\end{equation}
		 \EBY{Specifically}, their entries $  u_{s} $ and $ v_{s} $ represent the number of \BM{$ +1 $s and $ -1$s}, respectively, that $ C_{s} $ sends to $ V_{j} $ on $ b_{s,j} -\delta_{si}  $ of the $ b_{s,j} $ edges connecting $ C_{s} $ to $ V_{j}$. Thus, for \BM{$s = 1, 2, \ldots, m_{0}$} \EBY{we have}
		$0 \leqslant u_{s}\leqslant b_{s,j} -\delta_{si}$
		and $0 \leqslant v_{s}\leqslant b_{s,j} - \delta_{si}  -u_{s}$.}

\medskip

		\textbf{A posteriori update}
		
		For $ j= 1, 2, \ldots, n_{0} $
		if $ V_{j} $ is punctured \EBY{then}
		\begin{equation}
		 P_{\tapp}^{(\ell)}(j)= 
		\Pr\left\{  \tilde{L}^{(\ell)}_{\tin} \leqslant 0  \right\}.
		\end{equation}
		Otherwise, \EBY{we have} 
		\begin{equation}
	 P_{\tapp}^{(\ell)}(j)= 
		\sum\limits_{z} \Pr\left\{  \tilde{L}^{(\ell)}_{\tin} = z \right\}  Q\left( \dfrac{z+\mu_{\tch}}{\sigma_{\tch}}\right).
		\end{equation}
		
		
		\GL{We have
		\begin{equation}\label{eq:tildeLinAlgS}
		\begin{aligned}
		\Pr\left\{  \tilde{L}^{(\ell)}_{\tin} = z \right\} &= \sum\limits_{\tilde{\mathbf{u}}, \tilde{\mathbf{v}}} \prod\limits_{b_{s,j}\neq 0} 
		{\textstyle \binom{b_{s,j} }{\tilde{u}_{s},\tilde{v}_{s},b_{s,j}  - \tilde{u}_{s} - \tilde{v}_{s}}}\times\\
		&\left( 1 - 	q^{(\ell)}_{0}(s,j)  -	q^{(\ell)}_{-1}(s,j) \right) ^{\tilde{u}_{s}}\times \\
		&q^{(\ell)}_{-1}(s,j)^{\tilde{v}_{s}} q^{(\ell)}_{0}(s,j)^{b_{s,j}  - \tilde{u}_{s} - \tilde{v}_{s}}
		\end{aligned}
		\end{equation}
		where the outer sum is over all integer vector pairs $\tilde{\mathbf{u}}, \tilde{\mathbf{v}}$  for which 
		\[
		 \sum\limits_{e=1}^{m_{0}}w^{(\ell)}_{e,j}(\tilde{u}_{e}-\tilde{v}_{e}) = z 
		\]
		and
		where $	w^{(\ell)}_{e,j} $ is given in \eqref{eq:wkjAlgoS}. The vector elements $  \tilde{u}_{s} $ and $ \tilde{v}_{s} $ represent the number of \BM{$ +1 $s and $ -1$s}, respectively, that $ C_{s} $ sends to $ V_{j} $ on  the $ b_{s,j} $ edges connecting $ C_{s} $ to $ V_{j}$. Thus, for $s = 1, 2, \ldots, m_{0}$ \EBY{we have} 
		$0 \leqslant \tilde{u}_{s}\leqslant b_{s,j} $
		and $0 \leqslant \tilde{v}_{s}\leqslant b_{s,j} -\tilde{u}_{s}$.}
\end{enumerate}

\medskip

The protograph ensemble iterative decoding threshold $ (E_{\tb}/N_{0})^\star $ is defined as the minimum $ E_{\tb}/N_{0} $ for which $ P_{\tapp}^{(\ell)}(j) \to 0 $ for $ j=1, 2, \ldots, n_{0}$ as \GL{\BM{$ \ell \to \infty$}}.

\subsection{Stability Condition}
Stability analysis of \ac{DE} examines the convergence of the error and erasure probabilities to zero under the assumption that they are sufficiently small.
\GL{We will derive the stability condition for \ac{TMP} decoding for unstructured \ac{LDPC} ensembles only.} The derivation for protograph-based \ac{LDPC} ensembles is similar but more involved. We define $ \mathbf{p}^{(\ell)} $, the vector containing the erasure and error probabilities of the \ac{VN} messages and $ \mathbf{q}^{(\ell)} $, the vector containing the erasure and error probabilities of the \ac{CN} messages in the $ \ell$-th iteration as
\begin{align}
\mathbf{p}^{(\ell)} &= \left[  p_{0}^{(\ell)} , p_{-1}^{(\ell)} \right]^{T} & \text{and} & & \mathbf{q}^{(\ell)} &= \left[  q_{0}^{(\ell)} , q_{-1}^{(\ell)} \right]^{T}.
\end{align}

\EBY{Consider} first $ (d_{\tv},d_{\tc}) $ regular \GL{\ac{LDPC} code ensembles}. We should determine the evolution of $ \mathbf{p}^{(\ell)} $ over one iteration when we are close to the fixed point $ \mathbf{p}^{*}= \mathbf{0} $. Note that, as $ \mathbf{p} \to \mathbf{0} $, \EBY{we have} $ \mathbf{q} \to \mathbf{0} $ and as a result $ D \to +\infty $. Thus, for small \GL{error and erasure probabilities \EBY{we compute} 
\begin{align}
\begin{split}
p_{0}^{(\ell)} =&\alpha \Pr\left\{M_{\tin}^{(\ell)} = 0\right\}
\end{split}\\
p_{-1}^{(\ell)} =&  \beta \Pr\left\{M_{\tin}^{(\ell)} =0 \right\} + \Pr\left\{M_{\tin}^{(\ell)} \leqslant -1  \right\}
\end{align}}

\noindent
where, $ \Pr\left\{M_{\tin}^{(\ell)} = m_{\tin}\right\} $ can be calculated from \eqref{eq:PrMin} with $ d= d_{\tv}$\EBY{,} and for ease of notation we defined $ \alpha $ and $ \beta $ as 
\begin{align}
\alpha := \Pr\left\{-a \leqslant L_{\tch} \leqslant a\right\} & &\text{and} & &\beta := \Pr\left\{ L_{\tch} < -a\right\}.
\end{align}
\GL{Note that the parameters $\alpha,\beta$  summarize the role of the channel within the stability condition.}


\GL{We have 
\begin{equation}\label{eq:derivp0p0}
\begin{aligned}
 \lim\limits_{\mathbf{p}^{(\ell-1)} \to \mathbf{0}} {\dfrac{\partial p_{0}^{(\ell)}} {\partial p_{0}^{(\ell-1)}}} =\begin{cases}
 \alpha (d_{\tc}-1)  & d_{\tv} = 2 \\
  0 & \text{otherwise}.
 \end{cases}
 \end{aligned}
\end{equation}}
Similarly, \EBY{we have} 
\begin{align}
\lim\limits_{\mathbf{p}^{(\ell-1)} \to \mathbf{0}} {\dfrac{\partial p_{0}^{(\ell)}} {\partial p_{-1}^{(\ell-1)}}} =& \begin{cases}
2\alpha  (d_{\tc}-1)  & d_{\tv} = 3 \\
0 & \text{otherwise},
\end{cases} \label{eq:derivp0p-1}\\
\lim\limits_{\mathbf{p}^{(\ell-1)} \to \mathbf{0}} {\dfrac{\partial p_{-1}^{(\ell)}} {\partial p_{0}^{(\ell-1)}}} =& \begin{cases}
\beta   (d_{\tc}-1)  & d_{\tv} = 2 \\
0 & \text{otherwise}
\end{cases} \label{eq:derivp-1p0}
\end{align}
\begin{equation}\label{eq:derivp-1p-1}
\lim\limits_{\mathbf{p}^{(\ell-1)} \to \mathbf{0}} {\dfrac{\partial p_{-1}^{(\ell)}} {\partial p_{-1}^{(\ell-1)}}} = \begin{cases}
 (d_{\tc}-1)  & d_{\tv} = 2\\
2\beta   (d_{\tc}-1)  & d_{\tv} = 3 \\
0 & \text{otherwise}.
\end{cases}
\end{equation}
For unstructured \ac{LDPC} ensembles\EBY{,} \EBY{the first order Taylor expansions via}  \GL{\eqref{eq:derivp0p0}, \eqref{eq:derivp0p-1}, \eqref{eq:derivp-1p0} and \eqref{eq:derivp-1p-1}} yield
\begin{equation}
\mathbf{p}^{(\ell)}= 
\mathbf{J}\cdot \mathbf{p}^{(\ell-1)}
\end{equation}
where 
\begin{equation}
\mathbf{J}:= \rho'(1) \cdot \begin{bmatrix}
\alpha\lambda_{2} & 2\alpha\lambda_{3}  \\
\beta\lambda_{2} &  (\lambda_{2}+2\beta \lambda_{3})
\end{bmatrix}.
\end{equation}

\EBY{Let} $ \gamma $ \EBY{be} the spectral radius of $ \mathbf{J} $, i.e., the largest magnitude of its eigenvalues. We have
\begin{equation}
\begin{aligned}
\gamma=&  \frac{\rho'(1)}{2} \Bigg[\left(\alpha+1\right)\lambda_{2}+2\beta\lambda_{3}\Bigg.  \\
& + \Bigg. \sqrt{\left(\alpha-1\right)^{2}\lambda_{2}^{2}+4\beta^{2}\lambda_{3}^{2}+4\beta\lambda_{2}\lambda_{3}\left(\alpha+1\right) } \Bigg].
\end{aligned}
\end{equation}

\noindent
The stability condition is fulfilled if and only if $   \gamma < 1 $.

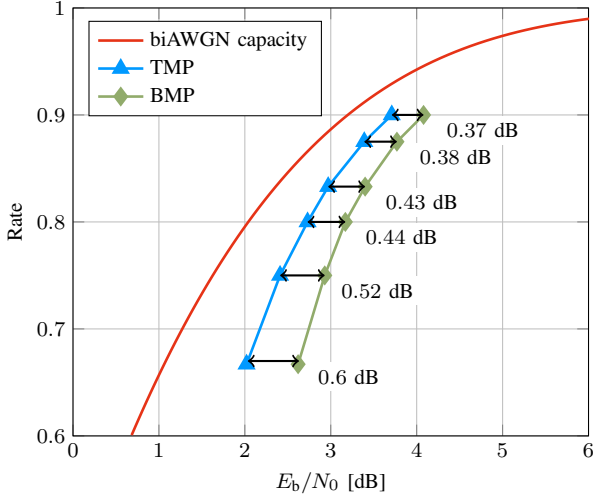
\begin{figure}[t]
    \centering
    \begin{tikzpicture}
\pgfplotsset{
  set layers,
  mark layer=axis tick labels
}
\begin{axis}
[
xmin=0,
xmax=6,
xlabel={$E_\tb/N_0$ [\si{dB}]},
ymin=0.6,
ymax=1,
grid=both,
ylabel={Rate},
legend pos=north west,
legend cell align=left,
]
\footnotesize

\path[name path global=line23] (0,0.67) -- (6,0.67);
\path[name path global=line34] (0,0.75) -- (6,0.75);
\path[name path global=line45] (0,0.80) -- (6,0.80);
\path[name path global=line56] (0,0.833) -- (6,0.833);
\path[name path global=line78] (0,0.875) -- (6,0.875);
\path[name path global=line910] (0,0.90) -- (6,0.90);

\addplot[line width=1,color = TUMBeamerRed, mark=none, mark options={solid}]
table[x=EbN0dB ,y=C,col sep=space,trim cells=true] {data/biAWGN-cap.txt};
\addlegendentry{biAWGN capacity};

\addplot[name path global=tmp,line width=1,TUMBeamerBlue,mark=triangle*,mark options={scale=1.5,fill=TUMBeamerBlue}]
table[x=threshold ,y=R,col sep=space,trim cells=true] {data/TMP-thresholds.txt};
\addlegendentry{TMP};

\addplot[name path global=bmp,line width=1,TUMBeamerGreen,mark=diamond*,mark options={scale=1.5,fill=TUMBeamerGreen}]
table[x=threshold ,y=R,col sep=space,trim cells=true] {data/BMP-thresholds.txt};
\addlegendentry{BMP};

\path[name intersections={of=line23 and bmp, name=p1}, name intersections={of=line23 and tmp, name=p2}];
\draw[<->,thick] let \p1=(p2-1), \p2=(p1-1) in (p1-1) -- (p2-1) node [below,midway,fill=white,yshift=0.0cm,xshift=1cm] {%
        \pgfplotsconvertunittocoordinate{x}{\x1}%
        \pgfplotscoordmath{x}{datascaletrafo inverse to fixed}{\pgfmathresult}%
        \edef\valueA{\pgfmathresult}%
        \pgfplotsconvertunittocoordinate{x}{\x2}%
        \pgfplotscoordmath{x}{datascaletrafo inverse to fixed}{\pgfmathresult}%
        \pgfmathparse{\pgfmathresult - \valueA}%
        \pgfmathprintnumber{\pgfmathresult} dB
};

\path[name intersections={of=line34 and bmp, name=p1}, name intersections={of=line34 and tmp, name=p2}];
\draw[<->,thick] let \p1=(p2-1), \p2=(p1-1) in (p1-1) -- (p2-1) node [below,midway,fill=white,yshift=0.0cm,xshift=1cm] {%
        \pgfplotsconvertunittocoordinate{x}{\x1}%
        \pgfplotscoordmath{x}{datascaletrafo inverse to fixed}{\pgfmathresult}%
        \edef\valueA{\pgfmathresult}%
        \pgfplotsconvertunittocoordinate{x}{\x2}%
        \pgfplotscoordmath{x}{datascaletrafo inverse to fixed}{\pgfmathresult}%
        \pgfmathparse{\pgfmathresult - \valueA}%
        \pgfmathprintnumber{\pgfmathresult} dB
};

\path[name intersections={of=line45 and bmp, name=p1}, name intersections={of=line45 and tmp, name=p2}];
\draw[<->,thick] let \p1=(p2-1), \p2=(p1-1) in (p1-1) -- (p2-1) node [below,midway,fill=white,yshift=0.0cm,xshift=1cm] {%
        \pgfplotsconvertunittocoordinate{x}{\x1}%
        \pgfplotscoordmath{x}{datascaletrafo inverse to fixed}{\pgfmathresult}%
        \edef\valueA{\pgfmathresult}%
        \pgfplotsconvertunittocoordinate{x}{\x2}%
        \pgfplotscoordmath{x}{datascaletrafo inverse to fixed}{\pgfmathresult}%
        \pgfmathparse{\pgfmathresult - \valueA}%
        \pgfmathprintnumber{\pgfmathresult} dB
};

\path[name intersections={of=line56 and bmp, name=p1}, name intersections={of=line56 and tmp, name=p2}];
\draw[<->,thick] let \p1=(p2-1), \p2=(p1-1) in (p1-1) -- (p2-1) node [below,midway,fill=white,yshift=0.0cm,xshift=1cm] {%
        \pgfplotsconvertunittocoordinate{x}{\x1}%
        \pgfplotscoordmath{x}{datascaletrafo inverse to fixed}{\pgfmathresult}%
        \edef\valueA{\pgfmathresult}%
        \pgfplotsconvertunittocoordinate{x}{\x2}%
        \pgfplotscoordmath{x}{datascaletrafo inverse to fixed}{\pgfmathresult}%
        \pgfmathparse{\pgfmathresult - \valueA}%
        \pgfmathprintnumber{\pgfmathresult} dB
};

\path[name intersections={of=line78 and bmp, name=p1}, name intersections={of=line78 and tmp, name=p2}];
\draw[<->,thick] let \p1=(p2-1), \p2=(p1-1) in (p1-1) -- (p2-1) node [below,midway,fill=white,yshift=0.0cm,xshift=1cm] {%
        \pgfplotsconvertunittocoordinate{x}{\x1}%
        \pgfplotscoordmath{x}{datascaletrafo inverse to fixed}{\pgfmathresult}%
        \edef\valueA{\pgfmathresult}%
        \pgfplotsconvertunittocoordinate{x}{\x2}%
        \pgfplotscoordmath{x}{datascaletrafo inverse to fixed}{\pgfmathresult}%
        \pgfmathparse{\pgfmathresult - \valueA}%
        \pgfmathprintnumber{\pgfmathresult} dB
};

\path[name intersections={of=line910 and bmp, name=p1}, name intersections={of=line910 and tmp, name=p2}];
\draw[<->,thick] let \p1=(p2-1), \p2=(p1-1) in (p1-1) -- (p2-1) node [below,midway,fill=white,yshift=0.0cm,xshift=1cm] {%
        \pgfplotsconvertunittocoordinate{x}{\x1}%
        \pgfplotscoordmath{x}{datascaletrafo inverse to fixed}{\pgfmathresult}%
        \edef\valueA{\pgfmathresult}%
        \pgfplotsconvertunittocoordinate{x}{\x2}%
        \pgfplotscoordmath{x}{datascaletrafo inverse to fixed}{\pgfmathresult}%
        \pgfmathparse{\pgfmathresult - \valueA}%
        \pgfmathprintnumber{\pgfmathresult} dB
};

	\end{axis}
\end{tikzpicture}%
    \caption{\BM{Decoding thresholds under BMP and TMP decoding} of optimized \GL{protograph \ac{LDPC} code ensembles.}}
    \label{fig:BMP-TMP-thresholds}
\end{figure}
%

\section{Code Design and Results}\label{sec:results}
\label{sec:codedesignresults}

\ac{DE} provides a criterion to design protographs with good waterfall performance. \EBY{In order to control} the error floor, 
we put constraints on the weight spectral shape $ G(\omega)$ of an \ac{LDPC} code ensemble, where
\begin{equation}
G(\omega):= \lim\limits_{n\to \infty }\dfrac{1}{n} \log_{2} \left( \mathcal{A}_{\omega n}\right).
\end{equation}
\EBY{Here}, $ \cA_{\omega n} $ is the expected number of weight $ \omega n$ codewords for an \ac{LDPC} code drawn randomly from the ensemble. Let $ \omega^\star =  \inf \{ \omega > 0 | G(\omega) = 0 \} $. If $ \omega^\star $ exists and $ G(\omega) < 0 $ for $ 0 < \omega < \omega^\star $ then $ \omega^\star $ is called the typical relative minimum distance \cite{weightenumerator_divsalar_2006}. We \BM{constrain} the ensemble search to ensembles with a strictly positive typical minimum distance. An efficient method to compute $ G(\omega)$ for protograph ensembles is presented in \cite{paolini_growthrate_2016}.

To find optimized protograph ensembles we apply differential evolution \cite{storn_differential_1997} and impose the above-mentioned  constraint. We use the decoding threshold as the cost function. The stability condition is used to discard ensembles at a preliminary stage. \BM{Due to space limitations a complete set of base matrices and decoder parameters $D^{(\ell) }_{CV}$ and $a$ are provided in \cite{BenYacoub2018arxiv}}.

\subsection{Numerical Results}\label{subsec:numericalresults}
\begin{figure}[t]
    \centering
    \footnotesize
    \begin{tikzpicture}
\footnotesize

\begin{axis}[
xlabel={$E_\tb/N_0$ [\si{dB}]},
ylabel={FER},
grid=both,
legend pos=south west,
legend cell align={left},
ymode=log,
ymin=1e-5,
]

\addplot[x filter/.code={\pgfmathparse{\pgfmathresult-1.76}\pgfmathresult},line width=1,TUMBeamerGreen,mark=diamond*, mark options={fill=TUMBeamerGreen}] table[x=snr,y=fer] {data/results-tmp-R=0.75-iter=30.txt};\label{plt:tmp_34}
\addplot[x filter/.code={\pgfmathparse{\pgfmathresult-2.218}\pgfmathresult},line width=1,TUMBeamerOrange,mark=triangle*, mark options={fill=TUMBeamerOrange}] table[x=snr,y=fer] {data/results-tmp-R=0.83-iter=30.txt};\label{plt:tmp_56}
\addplot[x filter/.code={\pgfmathparse{\pgfmathresult-2.43}\pgfmathresult},line width=1,TUMBeamerRed,mark=pentagon*, mark options={fill=TUMBeamerRed}] table[x=snr,y=fer] {data/results-tmp-R=0.87-iter=30.txt};\label{plt:tmp_78}

\addplot[x filter/.code={\pgfmathparse{\pgfmathresult-1.76}\pgfmathresult},line width=1,TUMBeamerGreen,mark=diamond*,mark options={solid, fill=TUMBeamerGreen},dashed] table[x=snr,y=fer] {data/results-ar4ja-tmp-R=0.75-iter=30.txt};\label{plt:ar4ja_tmp_34}
\addplot[x filter/.code={\pgfmathparse{\pgfmathresult-2.218}\pgfmathresult},line width=1,TUMBeamerOrange,mark=triangle*,mark options={solid, fill=TUMBeamerOrange},dashed] table[x=snr,y=fer] {data/results-ar4ja-tmp-R=0.83-iter=30.txt};\label{plt:ar4ja_tmp_56}
\addplot[x filter/.code={\pgfmathparse{\pgfmathresult-2.43}\pgfmathresult},line width=1,TUMBeamerRed,mark=pentagon*,mark options={solid,fill=TUMBeamerRed},dashed] table[x=snr,y=fer] {data/results-ar4ja-tmp-R=0.87-iter=30.txt};\label{plt:ar4ja_tmp_78}

\addplot[x filter/.code={\pgfmathparse{\pgfmathresult-1.76}\pgfmathresult},line width=1,TUMBeamerGreen,mark=diamond*,mark options={solid,fill=TUMBeamerGreen},dotted] table[x=snr,y=fer] {data/results-ar4ja-R=0.75-iter=30.txt};\label{plt:full_34}
\addplot[x filter/.code={\pgfmathparse{\pgfmathresult-2.218}\pgfmathresult},line width=1,TUMBeamerOrange,mark=triangle*,mark options={solid,fill=TUMBeamerOrange},dotted] table[x=snr,y=fer] {data/results-ar4ja-R=0.83-iter=30.txt};\label{plt:full_56}
\addplot[x filter/.code={\pgfmathparse{\pgfmathresult-2.43}\pgfmathresult},line width=1,TUMBeamerRed,mark=pentagon*,mark options={solid,fill=TUMBeamerRed},dotted] table[x=snr,y=fer] {data/results-ar4ja-R=0.87-iter=30.txt};\label{plt:full_78}



\end{axis}
\end{tikzpicture}
    \caption{FER versus $E_{\tb}/N_0$ for TMP and \GL{unquantized} BP decoding for  $R = 3/4$ (\protect\tikz[baseline=-0.5ex]{\protect\draw[line width=1,TUMBeamerGreen] (0,0) -- (.5,0)}), $R = 5/6$ (\protect\tikz[baseline=-0.5ex]{\protect\draw[line width=1,TUMBeamerOrange] (0,0) -- (.5,0)}) and $R = 7/8$ (\protect\tikz[baseline=-0.5ex]{\protect\draw[line width=1,TUMBeamerRed] (0,0) -- (.5,0)}). We compare the TMP performance of optimized codes (\ref{plt:tmp_34}, \ref{plt:tmp_56}, \ref{plt:tmp_78}) to their AR4JA counterparts with unquantized \ac{BP} (\ref{plt:full_34}, \ref{plt:full_56}, \ref{plt:full_78}) and TMP decoding (\ref{plt:ar4ja_tmp_34}, \ref{plt:ar4ja_tmp_56}, \ref{plt:ar4ja_tmp_78}).}
    \label{fig:Coded Results of TMP}
\end{figure}
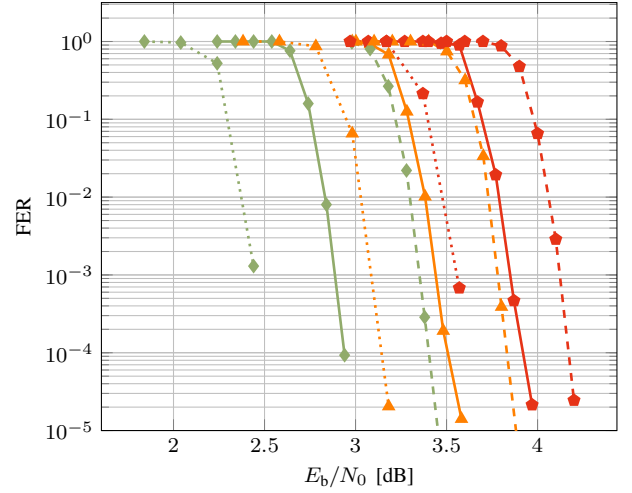

First, we investigate the gains of \ac{TMP} over \GL{\ac{BMP}} in terms of the \BM{iterative} decoding threshold. For both algorithms we \BM{obtain individually} optimized protograph ensembles for rates \EBY{$R\in\{2/3, 3/4, 4/5, 5/6, 7/8, 9/10\}$}, where we \GL{restrict} the maximum \ac{VN} degree to $ 20$ and $ L_{\tmax} = 200$. For \ac{BMP}, we simply replace the \BM{$ f$}-function  in \eqref{eq:VNupdatel0} and \eqref{eq:VNupdateAlgoS1} by the $ \sign $ function. \BM{Observe from Fig.~\ref{fig:BMP-TMP-thresholds}} that the gap to the Shannon limit decreases as \GL{the rate} increases. \ac{TMP} decoding \EBY{improves} \ac{BMP} decoding in particular for lower code rates. \BM{For $R = 2/3$,} the decoding threshold improves by \SI{0.6}{dB}  compared to \ac{BMP}, \BM{while} for $R = 9/10$, the gain is \SI{0.37}{dB}. 

\BM{To check the finite-length performance under \ac{TMP}, we design a further set of optimized protograph ensembles with rates} $ R \in \{3/4,4/5,5/6,7/8 \}$. \BM{To reduce decoding complexity}, we limit the maximum \ac{VN} degree to $ 12$ and the maximum number of decoding iterations \EBY{to} $ L_{\tmax} = 30$. All codes have a block length of \num{22176}, a quasi-cyclic structure, and are obtained by lifting \BM{the protographs by} a circulant version of the \ac{PEG} algorithm \cite{hu_regular_2005}.
Simulation results for rates 3/4, 5/6 and 7/8 are shown in Fig. \ref{fig:Coded Results of TMP} in terms of \ac{FER} \BM{versus $E_{\tb}/N_0$}. As a reference, we compare with \GL{\ac{AR4JA}} codes designed for unquantized \ac{BP} decoding \cite{divsalar_capacity-approaching_2009}. \GL{Observe that the protograph codes optimized for \ac{TMP} perform, under \ac{TMP} decoding,  remarkably close to \ac{AR4JA} codes decoded with unquantized \ac{BP}. The loss is limited to \SI{0.5}{dB} for the case of $R=3/4$, and reduces to \SI{0.2}{dB} for  $R=7/8$. When \ac{TMP} is used to decode the \ac{AR4JA} codes, the protograph \ac{LDPC} codes optimized for \ac{TMP} decoding outperform \ac{AR4JA} codes by \SI{0.5}{dB} at $R=3/4$ and by \SI{0.2}{dB}  at $R=7/8$.} 

\begin{figure}[t]
    \centering
    \footnotesize
    \begin{tikzpicture}

\begin{axis}
[
xmin=0,
xmax=0.1,
xlabel={$ \omega $},
ymin=-0.03,
grid=both,
ylabel={$ G(\omega) $},
legend pos=south east,
legend cell align=left,
]

\addplot[line width=1,TUMBeamerRed,mark=none, mark options={solid}]
table[x=omega ,y=G,col sep=space,trim cells=true] {data/growthrate4_5.txt};
\addlegendentry{TMP optimized, $ R = 4/5 $};

\addplot[line width=1,TUMBeamerBlue,mark=none, mark options={solid}]
table[x=omega ,y=G,col sep=space,trim cells=true] {data/growthrateARJA4_5.txt};
\addlegendentry{AR4JA, $R = 4/5$};

\addplot[line width=1,TUMBeamerGreen,mark=none, mark options={solid}]
table[x=omega ,y=G,col sep=space,trim cells=true] {data/growthraterandom4_5.txt};
\addlegendentry{Random codes};

\addplot[color = black,dashed,mark=none, mark options={solid}]
	table[x=omega,y=G,col sep=space,trim cells=true] {data/line0.txt};

\coordinate (A) at (axis cs: 3e-2,0.22);

	\end{axis}
	
\node[fill=white] at (A) {\input{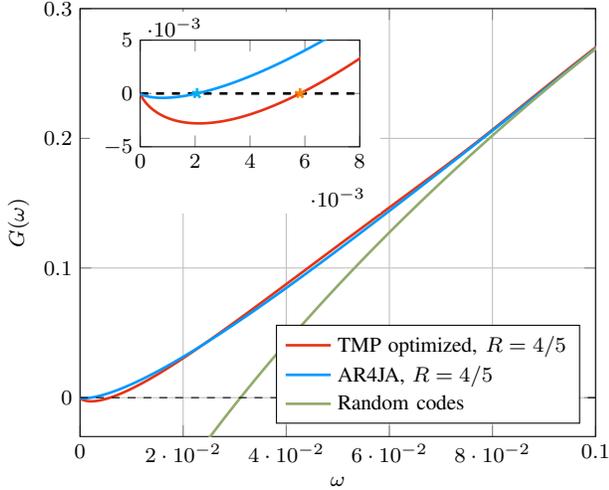}};	
\end{tikzpicture}%
    \caption{Weight spectral shape for \GL{rate-$4/5$ TMP-optimized protograph ensemble, an \ac{AR4JA} ensemble and the random code ensemble}.}
    \label{fig:growthrate}
\end{figure}

Typical relative minimum distances  of the protograph ensembles are given in Tab.~\ref{tab:5:omega}. The weight spectral shape for the optimized code \GL{ensemble and the AR4JA ensemble} with $ R = 4/5 $ is shown in Fig.~\ref{fig:growthrate}. 

%
\begin{table}[t]
	\begin{center}
		\caption{$\omega^\star$ for various protograph ensembles.}
		\label{tab:5:omega}
		\begin{tabular}{ccc}
			\thickhline  
	             $R$ & Design for TMP & AR4JA ensemble\\  \hline
	3/4& $0.00911$& $0.003227$ \\	
	4/5& $ 0.005824$& $ 0.002072$\\
	5/6& $0.003747$& $0.0014518$ \\
	7/8& $0.0022195$& $ 0.0008342$ 	\\
	\thickhline
		\end{tabular}
	\end{center}
\end{table}

\section{Conclusion}\label{sec:conclusions}

\GL{A low-complexity \ac{TMP} decoding algorithm for unstructured and protograph-based \ac{LDPC} codes \EBY{was} introduced. An exact \ac{DE} analysis \EBY{was} developed, which allows to tune the reliability parameters of the decoding algorithm\EBY{,} and to derive the asymptotic iterative decoding threshold under \ac{TMP} decoding. The analysis \EBY{was} complemented by \EBY{deriving} of the stability condition. A design methodology based on \ac{TMP} \ac{DE} analysis and on the distance properties of the code ensembles \EBY{was} presented and used \EBY{to} construct protograph-based \ac{LDPC} ensembles that perform close to the theoretical limits in the high code rate regime.} \BM{Under \ac{TMP} decoding, our codes outperform standard protograph-based codes designed for the unquantized \ac{BP} algorithm.}

\bibliographystyle{IEEEtran}


\onecolumn
\clearpage
\appendices

\section{Designed Protograph Ensembles}

\def\arraystretch{2}

\begin{table}[h]
 \caption{Base matrices for the protograph-based \ac{LDPC} ensembles in Fig.~1}
\label{tab:base_matrices_Fig1}
\begin{center}
\begin{tabular}{ccc}
\hline\hline
Rate & Decoding algorithm & Base matrix \\ \hline
$2/3$ & TMP & $\left( 
\begin{array}{cccccc}
3&4&3&7&3&1\\
0&0&1&8&1&3\\
\end{array}
\right)$\\
$2/3$ & BMP & $\left( 
\begin{array}{cccccc}
3&5&3&3&9&4\\
0&3&1&1&9&0\\
\end{array}
\right)$\\
$3/4$ & TMP & $\left( 
\begin{array}{cccccccc}
3&5&4&10&3&4&7&3\\
0&11&0&3&1&0&3&1\\
\end{array}
\right)$\\
$3/4$ & BMP & $\left( 
\begin{array}{cccccccc}
3&1&3&9&4&7&4&3\\
0&4&1&3&0&11&0&1\\
\end{array}
\right)$\\
$4/5$ & TMP & $\left( 
\begin{array}{cccccccccc}
3&3&4&4&4&4&10&4&3&2\\
0&3&1&4&0&0&3&0&1&13\\
\end{array}
\right)$\\
$4/5$ & BMP & $\left( 
\begin{array}{cccccccccc}
3&4&4&4&13&4&2&13&3&4\\
0&0&0&2&4&1&16&3&2&0\\
\end{array}
\right)$\\
$5/6$ & TMP & $\left( 
\begin{array}{cccccccccccc}
3&4&5&4&4&4&6&3&4&2&4&2\\
0&0&3&0&0&16&4&1&0&2&2&4\\
\end{array}
\right)$\\
$5/6$ & BMP & $\left( 
\begin{array}{cccccccccccc}
3&9&9&6&0&5&4&4&5&4&4&4\\
0&3&3&5&15&2&0&2&1&0&0&0\\
\end{array}
\right)$\\
$7/8$ & TMP & $\left( 
\begin{array}{cccccccccccccccc}
3&5&4&4&3&1&4&5&1&7&4&4&0&4&4&3\\
0&3&0&0&4&3&0&2&3&3&6&0&18&0&0&1\\
\end{array}
\right)$\\
$7/8$ & BMP & $\left( 
\begin{array}{cccccccccccccccc}
3&3&4&4&4&6&4&4&6&1&4&4&4&4&0&6\\
0&6&1&2&1&3&0&0&3&4&0&0&1&0&20&6\\
\end{array}
\right)$\\
$9/10$ & TMP & $\left( 
\begin{array}{cccccccccccccccccccc}
3&3&4&4&6&4&4&0&4&2&0 &2&3&4&4&4&4&5&4&6\\
0&7&1&0&1&0&0&16&0&2&5&2&3&0&2&4&0&2&0&5\\
\end{array}
\right)$\\ 
$9/10$ & BMP & $\left( 
\begin{array}{cccccccccccccccccccc}
3&0&4&3&5&4&4&9&4&4&6&4&8&5&3&3&4&4&3&3\\
0&4&0&1&3&0&16&1&0&0&2&9&6&2&3&4&0&0&1&1\\
\end{array}
\right)$\\ \hline\hline
\end{tabular}
\end{center}
\end{table}

\setcounter{MaxMatrixCols}{60}

\begin{table}[h]
 \caption{Base matrices for the protograph-based \ac{LDPC} ensembles in Fig.~2/3}
\label{tab:base_matrices_Fig2}
\begin{center}
\begin{tabular}{ccc}
\hline\hline
Rate & Decoding algorithm & Base matrix  \\ \hline
3/4 & TMP & $\vect{2 &3 &1 &4 &3 &5 &4 &3\\1 &1 &7 &0 &1 &6 &0 &1}$\\
4/5 & TMP & $\vect{2 &4 &2 &3 &4 &4 &4 &4 &2 &2\\ 1 &6 &2 &1 &0 &0 &1 &0 &10 &2}$\\
5/6 & TMP & $\vect{2 &2 &4 &2 &1 &3 &3 &4 &4 &2 &4 &1\\ 1 &2 &0 &2 &3 &9 &1 &0 &0 &2 &0 &11}$\\
7/8 & TMP & $\vect{2 &4 &4 &1 &4 &4 &4 &4 &4 &4 &4 &4 &4 &3 &2 &2\\1 &0 &8 &11 &0 &1 &1 &6 &1 &3 &0 &0 &0 &1 &4 &2}$\\ \hline\hline
\end{tabular}
\end{center}
\end{table}

\newpage

\EBY{As mentioned before, we choose the quantization parameter $ a $ to minimize the iterative decoding threshold of the considered protograph ensemble. For all protograph-based codes of Fig.~2, we get $a=1.3$.} In the following, we also provide the decoding weights $D_{CV}^{(\ell)}$ as explained in \eqref{eq:DCV} for these protograph-based codes. The format is \EBY{as follows}:

\begin{itemize}
\item Each row in the matrix refers to one BP iteration.
\item The entries of row $\ell$ comprise $m_0n_0$ numbers which denote the decoding weights  $D_{C_jV_i}^{(\ell)}, j = 1,\ldots, m_0, i = 1, \ldots, n_0$, such that $ D_{C_jV_i}^{(\ell)} = m_{\ell,(j-1)\cdot m_0 + i}$, where $m_{\ell,(j-1)\cdot m_0 + i}$ is the matrix element in the $\ell$-th row and $(j-1)\cdot m_0 + i$ position.
\end{itemize}

\begin{table*}[h]
\caption{Decoding weights $D_{CV}^{(\ell)}$ for TMP for rate 3/4 code of Fig. 2.}
\label{tab:decoding_weights_tmp34}

\[
{\renewcommand*{\arraystretch}{0.7}\vect{0.72 & 0.72 & 0.72 & 0.72 & 0.72 & 0.72 & 0.72 & 0.72 & 1.08 & 1.08 & 1.08 & 0.00 & 1.08 & 1.08 & 0.00 & 1.08 \\ 0.81 & 0.81 & 0.81 & 0.81 & 0.81 & 0.81 & 0.81 & 0.81 & 1.29 & 1.28 & 1.27 & 0.00 & 1.28 & 1.27 & 0.00 & 1.28 \\ 0.87 & 0.87 & 0.86 & 0.87 & 0.87 & 0.86 & 0.87 & 0.87 & 1.42 & 1.42 & 1.40 & 0.00 & 1.42 & 1.40 & 0.00 & 1.42 \\ 0.92 & 0.92 & 0.90 & 0.92 & 0.92 & 0.90 & 0.92 & 0.92 & 1.54 & 1.54 & 1.50 & 0.00 & 1.54 & 1.50 & 0.00 & 1.54 \\ 0.96 & 0.96 & 0.94 & 0.97 & 0.96 & 0.94 & 0.97 & 0.96 & 1.65 & 1.64 & 1.59 & 0.00 & 1.64 & 1.60 & 0.00 & 1.64 \\ 1.00 & 1.00 & 0.98 & 1.01 & 1.00 & 0.98 & 1.01 & 1.00 & 1.75 & 1.75 & 1.69 & 0.00 & 1.75 & 1.69 & 0.00 & 1.75 \\ 1.05 & 1.04 & 1.02 & 1.05 & 1.04 & 1.02 & 1.05 & 1.04 & 1.86 & 1.85 & 1.78 & 0.00 & 1.85 & 1.78 & 0.00 & 1.85 \\ 1.09 & 1.09 & 1.06 & 1.10 & 1.09 & 1.06 & 1.10 & 1.09 & 1.97 & 1.96 & 1.87 & 0.00 & 1.96 & 1.87 & 0.00 & 1.96 \\ 1.14 & 1.14 & 1.11 & 1.15 & 1.14 & 1.11 & 1.15 & 1.14 & 2.08 & 2.07 & 1.97 & 0.00 & 2.07 & 1.97 & 0.00 & 2.07 \\ 1.19 & 1.19 & 1.16 & 1.20 & 1.19 & 1.16 & 1.20 & 1.19 & 2.20 & 2.19 & 2.07 & 0.00 & 2.19 & 2.07 & 0.00 & 2.19 \\ 1.25 & 1.24 & 1.21 & 1.26 & 1.24 & 1.21 & 1.26 & 1.24 & 2.32 & 2.31 & 2.17 & 0.00 & 2.31 & 2.17 & 0.00 & 2.31 \\ 1.31 & 1.30 & 1.27 & 1.32 & 1.30 & 1.27 & 1.32 & 1.30 & 2.44 & 2.42 & 2.27 & 0.00 & 2.42 & 2.27 & 0.00 & 2.42 \\ 1.37 & 1.37 & 1.33 & 1.39 & 1.37 & 1.33 & 1.39 & 1.37 & 2.55 & 2.53 & 2.36 & 0.00 & 2.53 & 2.36 & 0.00 & 2.53 \\ 1.43 & 1.43 & 1.39 & 1.45 & 1.43 & 1.39 & 1.45 & 1.43 & 2.66 & 2.64 & 2.45 & 0.00 & 2.64 & 2.45 & 0.00 & 2.64 \\ 1.50 & 1.50 & 1.46 & 1.52 & 1.50 & 1.46 & 1.52 & 1.50 & 2.77 & 2.74 & 2.54 & 0.00 & 2.74 & 2.54 & 0.00 & 2.74 \\ 1.57 & 1.57 & 1.53 & 1.59 & 1.57 & 1.53 & 1.59 & 1.57 & 2.87 & 2.83 & 2.62 & 0.00 & 2.83 & 2.62 & 0.00 & 2.83 \\ 1.65 & 1.64 & 1.61 & 1.67 & 1.64 & 1.61 & 1.67 & 1.64 & 2.97 & 2.93 & 2.70 & 0.00 & 2.93 & 2.70 & 0.00 & 2.93 \\ 1.73 & 1.72 & 1.69 & 1.75 & 1.72 & 1.69 & 1.75 & 1.72 & 3.07 & 3.02 & 2.79 & 0.00 & 3.02 & 2.79 & 0.00 & 3.02 \\ 1.82 & 1.82 & 1.78 & 1.85 & 1.82 & 1.78 & 1.85 & 1.82 & 3.18 & 3.12 & 2.89 & 0.00 & 3.12 & 2.89 & 0.00 & 3.12 \\ 1.93 & 1.92 & 1.89 & 1.96 & 1.92 & 1.89 & 1.96 & 1.92 & 3.30 & 3.23 & 2.99 & 0.00 & 3.23 & 2.99 & 0.00 & 3.23 \\ 2.06 & 2.05 & 2.02 & 2.09 & 2.05 & 2.02 & 2.09 & 2.05 & 3.45 & 3.36 & 3.12 & 0.00 & 3.36 & 3.12 & 0.00 & 3.36 \\ 2.23 & 2.22 & 2.18 & 2.25 & 2.22 & 2.18 & 2.25 & 2.22 & 3.63 & 3.52 & 3.28 & 0.00 & 3.52 & 3.28 & 0.00 & 3.52 \\ 2.46 & 2.44 & 2.41 & 2.48 & 2.44 & 2.40 & 2.48 & 2.44 & 3.86 & 3.73 & 3.48 & 0.00 & 3.73 & 3.48 & 0.00 & 3.73 \\ 2.78 & 2.76 & 2.72 & 2.80 & 2.76 & 2.72 & 2.80 & 2.76 & 4.20 & 4.02 & 3.78 & 0.00 & 4.02 & 3.78 & 0.00 & 4.02 \\ 3.27 & 3.24 & 3.21 & 3.28 & 3.24 & 3.21 & 3.28 & 3.24 & 4.70 & 4.45 & 4.22 & 0.00 & 4.45 & 4.22 & 0.00 & 4.45 \\ 4.05 & 4.02 & 3.98 & 4.05 & 4.02 & 3.98 & 4.05 & 4.02 & 5.51 & 5.13 & 4.93 & 0.00 & 5.13 & 4.93 & 0.00 & 5.13 \\ 5.37 & 5.30 & 5.27 & 5.33 & 5.30 & 5.27 & 5.33 & 5.30 & 6.87 & 6.26 & 6.10 & 0.00 & 6.26 & 6.10 & 0.00 & 6.26 \\ 7.53 & 7.39 & 7.36 & 7.41 & 7.39 & 7.36 & 7.41 & 7.39 & 9.18 & 8.07 & 7.96 & 0.00 & 8.07 & 7.96 & 0.00 & 8.07 \\ 10.82 & 10.46 & 10.44 & 10.47 & 10.46 & 10.44 & 10.47 & 10.46 & 12.96 & 10.64 & 10.61 & 0.00 & 10.64 & 10.61 & 0.00 & 10.64 \\ 15.04 & 14.42 & 14.42 & 14.42 & 14.42 & 14.42 & 14.42 & 14.42 & 18.74 & 14.04 & 14.04 & 0.00 & 14.04 & 14.04 & 0.00 & 14.04}}
\]
\end{table*}

\newpage

\begin{table*}[h]
\caption{Decoding weights $D_{CV}^{(\ell)}$ for TMP for rate 4/5 code of Fig. 3.}
\label{tab:decoding_weights_tmp45}
\rotatebox{90}{%
${\renewcommand*{\arraystretch}{0.7}\vect{0.74 & 0.74 & 0.74 & 0.74 & 0.74 & 0.74 & 0.74 & 0.74 & 0.74 & 0.74 & 1.02 & 1.02 & 1.02 & 1.02 & 0.00 & 0.00 & 1.02 & 0.00 & 1.02 & 1.02 \\ 0.82 & 0.82 & 0.82 & 0.82 & 0.82 & 0.82 & 0.82 & 0.82 & 0.82 & 0.82 & 1.22 & 1.21 & 1.22 & 1.22 & 0.00 & 0.00 & 1.22 & 0.00 & 1.21 & 1.22 \\ 0.88 & 0.87 & 0.87 & 0.88 & 0.88 & 0.88 & 0.88 & 0.88 & 0.86 & 0.87 & 1.37 & 1.35 & 1.36 & 1.37 & 0.00 & 0.00 & 1.37 & 0.00 & 1.34 & 1.36 \\ 0.92 & 0.91 & 0.92 & 0.92 & 0.92 & 0.92 & 0.92 & 0.92 & 0.91 & 0.92 & 1.50 & 1.47 & 1.49 & 1.50 & 0.00 & 0.00 & 1.49 & 0.00 & 1.46 & 1.49 \\ 0.97 & 0.95 & 0.96 & 0.97 & 0.97 & 0.97 & 0.97 & 0.97 & 0.95 & 0.96 & 1.62 & 1.59 & 1.61 & 1.62 & 0.00 & 0.00 & 1.62 & 0.00 & 1.57 & 1.61 \\ 1.01 & 1.00 & 1.01 & 1.01 & 1.02 & 1.02 & 1.01 & 1.02 & 0.99 & 1.01 & 1.75 & 1.70 & 1.74 & 1.75 & 0.00 & 0.00 & 1.75 & 0.00 & 1.69 & 1.74 \\ 1.06 & 1.04 & 1.06 & 1.06 & 1.07 & 1.07 & 1.06 & 1.07 & 1.04 & 1.06 & 1.89 & 1.83 & 1.87 & 1.89 & 0.00 & 0.00 & 1.88 & 0.00 & 1.81 & 1.87 \\ 1.12 & 1.09 & 1.11 & 1.12 & 1.12 & 1.12 & 1.12 & 1.12 & 1.09 & 1.11 & 2.03 & 1.95 & 2.01 & 2.03 & 0.00 & 0.00 & 2.02 & 0.00 & 1.94 & 2.01 \\ 1.18 & 1.15 & 1.17 & 1.18 & 1.19 & 1.19 & 1.17 & 1.19 & 1.15 & 1.17 & 2.18 & 2.08 & 2.15 & 2.17 & 0.00 & 0.00 & 2.17 & 0.00 & 2.06 & 2.15 \\ 1.24 & 1.21 & 1.23 & 1.24 & 1.25 & 1.25 & 1.24 & 1.25 & 1.21 & 1.23 & 2.32 & 2.20 & 2.28 & 2.31 & 0.00 & 0.00 & 2.31 & 0.00 & 2.19 & 2.28 \\ 1.31 & 1.28 & 1.29 & 1.30 & 1.32 & 1.32 & 1.30 & 1.32 & 1.27 & 1.29 & 2.46 & 2.32 & 2.40 & 2.44 & 0.00 & 0.00 & 2.43 & 0.00 & 2.30 & 2.40 \\ 1.37 & 1.34 & 1.36 & 1.37 & 1.39 & 1.39 & 1.37 & 1.39 & 1.34 & 1.36 & 2.58 & 2.42 & 2.51 & 2.56 & 0.00 & 0.00 & 2.55 & 0.00 & 2.41 & 2.51 \\ 1.44 & 1.41 & 1.42 & 1.43 & 1.45 & 1.45 & 1.43 & 1.45 & 1.40 & 1.42 & 2.68 & 2.51 & 2.60 & 2.67 & 0.00 & 0.00 & 2.65 & 0.00 & 2.50 & 2.60 \\ 1.50 & 1.47 & 1.49 & 1.50 & 1.52 & 1.52 & 1.50 & 1.52 & 1.47 & 1.49 & 2.78 & 2.60 & 2.69 & 2.76 & 0.00 & 0.00 & 2.74 & 0.00 & 2.59 & 2.69 \\ 1.57 & 1.54 & 1.56 & 1.57 & 1.59 & 1.59 & 1.56 & 1.59 & 1.54 & 1.56 & 2.88 & 2.68 & 2.78 & 2.85 & 0.00 & 0.00 & 2.83 & 0.00 & 2.67 & 2.78 \\ 1.64 & 1.61 & 1.62 & 1.64 & 1.66 & 1.66 & 1.63 & 1.66 & 1.61 & 1.62 & 2.97 & 2.76 & 2.86 & 2.94 & 0.00 & 0.00 & 2.91 & 0.00 & 2.75 & 2.86 \\ 1.71 & 1.68 & 1.70 & 1.71 & 1.73 & 1.73 & 1.71 & 1.73 & 1.68 & 1.70 & 3.06 & 2.84 & 2.94 & 3.02 & 0.00 & 0.00 & 2.99 & 0.00 & 2.83 & 2.94 \\ 1.80 & 1.76 & 1.78 & 1.79 & 1.82 & 1.82 & 1.79 & 1.82 & 1.76 & 1.78 & 3.16 & 2.93 & 3.03 & 3.12 & 0.00 & 0.00 & 3.08 & 0.00 & 2.92 & 3.03 \\ 1.89 & 1.86 & 1.87 & 1.89 & 1.91 & 1.91 & 1.88 & 1.91 & 1.86 & 1.87 & 3.27 & 3.02 & 3.13 & 3.22 & 0.00 & 0.00 & 3.18 & 0.00 & 3.02 & 3.13 \\ 2.01 & 1.97 & 1.99 & 2.00 & 2.03 & 2.03 & 1.99 & 2.03 & 1.97 & 1.99 & 3.40 & 3.14 & 3.24 & 3.34 & 0.00 & 0.00 & 3.29 & 0.00 & 3.13 & 3.24 \\ 2.15 & 2.11 & 2.13 & 2.14 & 2.17 & 2.17 & 2.14 & 2.17 & 2.11 & 2.13 & 3.56 & 3.28 & 3.38 & 3.48 & 0.00 & 0.00 & 3.43 & 0.00 & 3.28 & 3.38 \\ 2.34 & 2.30 & 2.32 & 2.33 & 2.36 & 2.36 & 2.32 & 2.36 & 2.30 & 2.32 & 3.77 & 3.46 & 3.57 & 3.67 & 0.00 & 0.00 & 3.61 & 0.00 & 3.46 & 3.57 \\ 2.61 & 2.57 & 2.59 & 2.60 & 2.63 & 2.63 & 2.59 & 2.63 & 2.57 & 2.59 & 4.06 & 3.72 & 3.82 & 3.93 & 0.00 & 0.00 & 3.85 & 0.00 & 3.72 & 3.82 \\ 3.02 & 2.97 & 2.99 & 3.00 & 3.03 & 3.03 & 2.99 & 3.03 & 2.97 & 2.99 & 4.49 & 4.10 & 4.20 & 4.31 & 0.00 & 0.00 & 4.22 & 0.00 & 4.10 & 4.20 \\ 3.67 & 3.62 & 3.64 & 3.65 & 3.68 & 3.68 & 3.64 & 3.68 & 3.62 & 3.64 & 5.17 & 4.70 & 4.79 & 4.90 & 0.00 & 0.00 & 4.79 & 0.00 & 4.70 & 4.79 \\ 4.78 & 4.71 & 4.72 & 4.74 & 4.77 & 4.77 & 4.72 & 4.77 & 4.71 & 4.72 & 6.31 & 5.69 & 5.77 & 5.86 & 0.00 & 0.00 & 5.74 & 0.00 & 5.69 & 5.77 \\ 6.66 & 6.54 & 6.56 & 6.57 & 6.59 & 6.59 & 6.55 & 6.59 & 6.54 & 6.56 & 8.24 & 7.31 & 7.38 & 7.44 & 0.00 & 0.00 & 7.34 & 0.00 & 7.31 & 7.38 \\ 9.68 & 9.41 & 9.43 & 9.43 & 9.45 & 9.45 & 9.41 & 9.45 & 9.41 & 9.43 & 11.46 & 9.78 & 9.82 & 9.84 & 0.00 & 0.00 & 9.79 & 0.00 & 9.78 & 9.82 \\ 13.78 & 13.22 & 13.22 & 13.23 & 13.23 & 13.23 & 13.22 & 13.23 & 13.22 & 13.22 & 16.56 & 13.08 & 13.09 & 13.09 & 0.00 & 0.00 & 13.08 & 0.00 & 13.08 & 13.09}}$}
\end{table*}

\newpage

\begin{table*}[h]
\caption{Decoding weights $D_{CV}^{(\ell)}$ for TMP for rate 5/6 code of Fig. 2}
\label{tab:decoding_weights_tmp56}
\rotatebox{90}{
${\renewcommand*{\arraystretch}{0.7}\vect{0.87 & 0.87 & 0.87 & 0.87 & 0.87 & 0.87 & 0.87 & 0.87 & 0.87 & 0.87 & 0.87 & 0.87 & 0.90 & 0.90 & 0.00 & 0.90 & 0.90 & 0.90 & 0.90 & 0.00 & 0.00 & 0.90 & 0.00 & 0.90 \\ 0.95 & 0.95 & 0.95 & 0.95 & 0.95 & 0.95 & 0.95 & 0.95 & 0.95 & 0.95 & 0.95 & 0.95 & 1.07 & 1.07 & 0.00 & 1.07 & 1.07 & 1.07 & 1.07 & 0.00 & 0.00 & 1.07 & 0.00 & 1.07 \\ 0.99 & 0.99 & 0.99 & 0.99 & 0.99 & 0.98 & 0.99 & 0.99 & 0.99 & 0.99 & 0.99 & 0.98 & 1.19 & 1.18 & 0.00 & 1.18 & 1.18 & 1.17 & 1.18 & 0.00 & 0.00 & 1.18 & 0.00 & 1.17 \\ 1.02 & 1.02 & 1.02 & 1.02 & 1.02 & 1.01 & 1.02 & 1.02 & 1.02 & 1.02 & 1.02 & 1.01 & 1.28 & 1.28 & 0.00 & 1.28 & 1.28 & 1.26 & 1.28 & 0.00 & 0.00 & 1.28 & 0.00 & 1.26 \\ 1.05 & 1.05 & 1.05 & 1.05 & 1.05 & 1.04 & 1.05 & 1.05 & 1.05 & 1.05 & 1.05 & 1.04 & 1.37 & 1.37 & 0.00 & 1.37 & 1.36 & 1.35 & 1.37 & 0.00 & 0.00 & 1.37 & 0.00 & 1.35 \\ 1.08 & 1.08 & 1.08 & 1.08 & 1.08 & 1.06 & 1.08 & 1.08 & 1.08 & 1.08 & 1.08 & 1.06 & 1.46 & 1.45 & 0.00 & 1.45 & 1.45 & 1.43 & 1.46 & 0.00 & 0.00 & 1.45 & 0.00 & 1.43 \\ 1.11 & 1.11 & 1.11 & 1.11 & 1.10 & 1.09 & 1.11 & 1.11 & 1.11 & 1.11 & 1.11 & 1.09 & 1.55 & 1.55 & 0.00 & 1.55 & 1.54 & 1.52 & 1.55 & 0.00 & 0.00 & 1.55 & 0.00 & 1.52 \\ 1.14 & 1.14 & 1.14 & 1.14 & 1.14 & 1.12 & 1.14 & 1.14 & 1.14 & 1.14 & 1.14 & 1.12 & 1.66 & 1.65 & 0.00 & 1.65 & 1.65 & 1.62 & 1.66 & 0.00 & 0.00 & 1.65 & 0.00 & 1.61 \\ 1.18 & 1.17 & 1.18 & 1.17 & 1.17 & 1.16 & 1.18 & 1.18 & 1.18 & 1.17 & 1.18 & 1.15 & 1.78 & 1.77 & 0.00 & 1.77 & 1.76 & 1.73 & 1.78 & 0.00 & 0.00 & 1.77 & 0.00 & 1.73 \\ 1.22 & 1.22 & 1.23 & 1.22 & 1.21 & 1.20 & 1.22 & 1.23 & 1.23 & 1.22 & 1.23 & 1.20 & 1.92 & 1.90 & 0.00 & 1.90 & 1.89 & 1.85 & 1.91 & 0.00 & 0.00 & 1.90 & 0.00 & 1.85 \\ 1.27 & 1.27 & 1.28 & 1.27 & 1.26 & 1.25 & 1.27 & 1.28 & 1.28 & 1.27 & 1.28 & 1.24 & 2.07 & 2.05 & 0.00 & 2.05 & 2.04 & 1.99 & 2.06 & 0.00 & 0.00 & 2.05 & 0.00 & 1.99 \\ 1.33 & 1.32 & 1.34 & 1.32 & 1.32 & 1.30 & 1.33 & 1.34 & 1.34 & 1.32 & 1.34 & 1.30 & 2.23 & 2.20 & 0.00 & 2.20 & 2.18 & 2.13 & 2.22 & 0.00 & 0.00 & 2.20 & 0.00 & 2.13 \\ 1.40 & 1.39 & 1.40 & 1.39 & 1.38 & 1.37 & 1.39 & 1.40 & 1.40 & 1.39 & 1.40 & 1.36 & 2.39 & 2.35 & 0.00 & 2.35 & 2.33 & 2.27 & 2.38 & 0.00 & 0.00 & 2.35 & 0.00 & 2.27 \\ 1.46 & 1.45 & 1.47 & 1.45 & 1.45 & 1.43 & 1.46 & 1.47 & 1.47 & 1.45 & 1.47 & 1.43 & 2.53 & 2.48 & 0.00 & 2.48 & 2.46 & 2.40 & 2.52 & 0.00 & 0.00 & 2.48 & 0.00 & 2.40 \\ 1.54 & 1.52 & 1.55 & 1.52 & 1.52 & 1.51 & 1.53 & 1.55 & 1.55 & 1.52 & 1.55 & 1.50 & 2.66 & 2.60 & 0.00 & 2.60 & 2.58 & 2.52 & 2.65 & 0.00 & 0.00 & 2.60 & 0.00 & 2.52 \\ 1.61 & 1.60 & 1.62 & 1.60 & 1.59 & 1.58 & 1.61 & 1.62 & 1.62 & 1.60 & 1.62 & 1.58 & 2.78 & 2.71 & 0.00 & 2.71 & 2.68 & 2.62 & 2.76 & 0.00 & 0.00 & 2.71 & 0.00 & 2.62 \\ 1.69 & 1.67 & 1.70 & 1.67 & 1.67 & 1.66 & 1.68 & 1.70 & 1.70 & 1.67 & 1.70 & 1.65 & 2.89 & 2.81 & 0.00 & 2.81 & 2.78 & 2.72 & 2.87 & 0.00 & 0.00 & 2.81 & 0.00 & 2.72 \\ 1.77 & 1.75 & 1.78 & 1.75 & 1.75 & 1.74 & 1.76 & 1.78 & 1.78 & 1.75 & 1.78 & 1.74 & 3.00 & 2.91 & 0.00 & 2.91 & 2.88 & 2.82 & 2.97 & 0.00 & 0.00 & 2.91 & 0.00 & 2.82 \\ 1.86 & 1.84 & 1.87 & 1.84 & 1.83 & 1.82 & 1.85 & 1.87 & 1.87 & 1.84 & 1.87 & 1.82 & 3.12 & 3.01 & 0.00 & 3.01 & 2.98 & 2.92 & 3.08 & 0.00 & 0.00 & 3.01 & 0.00 & 2.92 \\ 1.96 & 1.94 & 1.97 & 1.94 & 1.94 & 1.93 & 1.95 & 1.97 & 1.97 & 1.94 & 1.97 & 1.93 & 3.24 & 3.12 & 0.00 & 3.12 & 3.09 & 3.04 & 3.19 & 0.00 & 0.00 & 3.12 & 0.00 & 3.03 \\ 2.08 & 2.06 & 2.09 & 2.06 & 2.05 & 2.05 & 2.07 & 2.09 & 2.09 & 2.06 & 2.09 & 2.05 & 3.38 & 3.25 & 0.00 & 3.25 & 3.21 & 3.16 & 3.32 & 0.00 & 0.00 & 3.25 & 0.00 & 3.16 \\ 2.23 & 2.21 & 2.24 & 2.21 & 2.20 & 2.20 & 2.22 & 2.24 & 2.24 & 2.21 & 2.24 & 2.20 & 3.55 & 3.40 & 0.00 & 3.40 & 3.37 & 3.32 & 3.48 & 0.00 & 0.00 & 3.40 & 0.00 & 3.32 \\ 2.43 & 2.41 & 2.44 & 2.41 & 2.40 & 2.39 & 2.42 & 2.44 & 2.44 & 2.41 & 2.44 & 2.39 & 3.76 & 3.60 & 0.00 & 3.60 & 3.56 & 3.52 & 3.68 & 0.00 & 0.00 & 3.60 & 0.00 & 3.52 \\ 2.71 & 2.69 & 2.72 & 2.69 & 2.68 & 2.67 & 2.70 & 2.72 & 2.72 & 2.69 & 2.72 & 2.67 & 4.06 & 3.87 & 0.00 & 3.87 & 3.83 & 3.79 & 3.95 & 0.00 & 0.00 & 3.87 & 0.00 & 3.79 \\ 3.13 & 3.10 & 3.14 & 3.10 & 3.10 & 3.09 & 3.11 & 3.14 & 3.14 & 3.10 & 3.14 & 3.09 & 4.50 & 4.27 & 0.00 & 4.27 & 4.23 & 4.19 & 4.35 & 0.00 & 0.00 & 4.27 & 0.00 & 4.19 \\ 3.80 & 3.77 & 3.80 & 3.77 & 3.76 & 3.76 & 3.78 & 3.80 & 3.80 & 3.77 & 3.80 & 3.76 & 5.19 & 4.89 & 0.00 & 4.89 & 4.85 & 4.81 & 4.96 & 0.00 & 0.00 & 4.89 & 0.00 & 4.81 \\ 4.93 & 4.88 & 4.91 & 4.88 & 4.87 & 4.87 & 4.89 & 4.91 & 4.91 & 4.88 & 4.91 & 4.87 & 6.32 & 5.89 & 0.00 & 5.89 & 5.86 & 5.83 & 5.96 & 0.00 & 0.00 & 5.89 & 0.00 & 5.83 \\ 6.83 & 6.74 & 6.77 & 6.74 & 6.73 & 6.73 & 6.75 & 6.77 & 6.77 & 6.74 & 6.77 & 6.73 & 8.23 & 7.55 & 0.00 & 7.55 & 7.53 & 7.50 & 7.60 & 0.00 & 0.00 & 7.55 & 0.00 & 7.50 \\ 9.88 & 9.64 & 9.65 & 9.64 & 9.63 & 9.62 & 9.64 & 9.65 & 9.65 & 9.64 & 9.65 & 9.62 & 11.42 & 10.08 & 0.00 & 10.08 & 10.07 & 10.05 & 10.09 & 0.00 & 0.00 & 10.08 & 0.00 & 10.05 \\ 14.02 & 13.46 & 13.46 & 13.46 & 13.46 & 13.46 & 13.46 & 13.46 & 13.46 & 13.46 & 13.46 & 13.46 & 16.49 & 13.45 & 0.00 & 13.45 & 13.45 & 13.44 & 13.45 & 0.00 & 0.00 & 13.45 & 0.00 & 13.44}}$}
\end{table*}

\newpage

\begin{table*}[h]
\caption{Decoding weights $D_{CV}^{(\ell)}$ for TMP for rate 7/8 code of Fig. 2}
\label{tab:decoding_weights_tmp}

\scalebox{0.75}{\rotatebox{90}{%
{\renewcommand*{\arraystretch}{0.7}$\vect{0.68 & 0.68 & 0.68 & 0.68 & 0.68 & 0.68 & 0.68 & 0.68 & 0.68 & 0.68 & 0.68 & 0.68 & 0.68 & 0.68 & 0.68 & 0.68 & 0.97 & 0.00 & 0.97 & 0.97 & 0.00 & 0.97 & 0.97 & 0.97 & 0.97 & 0.97 & 0.00 & 0.00 & 0.00 & 0.97 & 0.97 & 0.97 \\ 0.75 & 0.75 & 0.75 & 0.75 & 0.75 & 0.75 & 0.75 & 0.75 & 0.75 & 0.75 & 0.75 & 0.75 & 0.75 & 0.75 & 0.75 & 0.75 & 1.15 & 0.00 & 1.14 & 1.14 & 0.00 & 1.15 & 1.15 & 1.15 & 1.15 & 1.15 & 0.00 & 0.00 & 0.00 & 1.15 & 1.15 & 1.15 \\ 0.80 & 0.80 & 0.79 & 0.79 & 0.80 & 0.80 & 0.80 & 0.79 & 0.80 & 0.79 & 0.80 & 0.80 & 0.80 & 0.80 & 0.79 & 0.80 & 1.27 & 0.00 & 1.26 & 1.26 & 0.00 & 1.27 & 1.27 & 1.26 & 1.27 & 1.27 & 0.00 & 0.00 & 0.00 & 1.27 & 1.27 & 1.27 \\ 0.83 & 0.83 & 0.82 & 0.82 & 0.83 & 0.83 & 0.83 & 0.83 & 0.83 & 0.83 & 0.83 & 0.83 & 0.83 & 0.83 & 0.83 & 0.83 & 1.37 & 0.00 & 1.36 & 1.35 & 0.00 & 1.37 & 1.37 & 1.36 & 1.37 & 1.36 & 0.00 & 0.00 & 0.00 & 1.37 & 1.36 & 1.37 \\ 0.86 & 0.86 & 0.85 & 0.85 & 0.86 & 0.86 & 0.86 & 0.86 & 0.86 & 0.86 & 0.86 & 0.86 & 0.86 & 0.86 & 0.86 & 0.86 & 1.46 & 0.00 & 1.44 & 1.44 & 0.00 & 1.46 & 1.46 & 1.45 & 1.46 & 1.46 & 0.00 & 0.00 & 0.00 & 1.46 & 1.45 & 1.46 \\ 0.90 & 0.90 & 0.89 & 0.88 & 0.90 & 0.90 & 0.90 & 0.89 & 0.90 & 0.89 & 0.90 & 0.90 & 0.90 & 0.90 & 0.89 & 0.89 & 1.56 & 0.00 & 1.53 & 1.53 & 0.00 & 1.56 & 1.56 & 1.54 & 1.56 & 1.55 & 0.00 & 0.00 & 0.00 & 1.56 & 1.55 & 1.55 \\ 0.93 & 0.93 & 0.92 & 0.92 & 0.93 & 0.93 & 0.93 & 0.92 & 0.93 & 0.92 & 0.93 & 0.93 & 0.93 & 0.93 & 0.92 & 0.93 & 1.66 & 0.00 & 1.63 & 1.62 & 0.00 & 1.66 & 1.66 & 1.63 & 1.66 & 1.65 & 0.00 & 0.00 & 0.00 & 1.66 & 1.64 & 1.65 \\ 0.97 & 0.97 & 0.95 & 0.95 & 0.97 & 0.97 & 0.97 & 0.96 & 0.97 & 0.96 & 0.97 & 0.97 & 0.97 & 0.97 & 0.96 & 0.96 & 1.77 & 0.00 & 1.73 & 1.73 & 0.00 & 1.77 & 1.77 & 1.74 & 1.77 & 1.75 & 0.00 & 0.00 & 0.00 & 1.77 & 1.75 & 1.76 \\ 1.01 & 1.01 & 1.00 & 0.99 & 1.01 & 1.01 & 1.01 & 1.00 & 1.01 & 1.00 & 1.01 & 1.01 & 1.01 & 1.01 & 1.00 & 1.01 & 1.89 & 0.00 & 1.85 & 1.84 & 0.00 & 1.89 & 1.89 & 1.85 & 1.89 & 1.87 & 0.00 & 0.00 & 0.00 & 1.89 & 1.86 & 1.88 \\ 1.06 & 1.06 & 1.04 & 1.04 & 1.06 & 1.06 & 1.06 & 1.05 & 1.06 & 1.05 & 1.06 & 1.06 & 1.06 & 1.06 & 1.05 & 1.06 & 2.03 & 0.00 & 1.97 & 1.97 & 0.00 & 2.03 & 2.03 & 1.98 & 2.03 & 2.00 & 0.00 & 0.00 & 0.00 & 2.03 & 1.99 & 2.02 \\ 1.12 & 1.12 & 1.10 & 1.10 & 1.12 & 1.11 & 1.11 & 1.10 & 1.11 & 1.11 & 1.12 & 1.12 & 1.12 & 1.12 & 1.10 & 1.11 & 2.18 & 0.00 & 2.11 & 2.11 & 0.00 & 2.18 & 2.18 & 2.12 & 2.18 & 2.14 & 0.00 & 0.00 & 0.00 & 2.18 & 2.14 & 2.16 \\ 1.18 & 1.19 & 1.16 & 1.16 & 1.19 & 1.18 & 1.18 & 1.16 & 1.18 & 1.17 & 1.19 & 1.19 & 1.19 & 1.18 & 1.17 & 1.17 & 2.34 & 0.00 & 2.26 & 2.25 & 0.00 & 2.33 & 2.33 & 2.27 & 2.33 & 2.29 & 0.00 & 0.00 & 0.00 & 2.34 & 2.28 & 2.31 \\ 1.25 & 1.26 & 1.23 & 1.23 & 1.26 & 1.25 & 1.25 & 1.23 & 1.25 & 1.24 & 1.26 & 1.26 & 1.26 & 1.25 & 1.24 & 1.24 & 2.50 & 0.00 & 2.40 & 2.40 & 0.00 & 2.49 & 2.49 & 2.41 & 2.49 & 2.44 & 0.00 & 0.00 & 0.00 & 2.50 & 2.43 & 2.47 \\ 1.32 & 1.33 & 1.30 & 1.30 & 1.33 & 1.32 & 1.32 & 1.31 & 1.32 & 1.31 & 1.33 & 1.33 & 1.33 & 1.32 & 1.31 & 1.32 & 2.66 & 0.00 & 2.54 & 2.54 & 0.00 & 2.64 & 2.64 & 2.55 & 2.64 & 2.57 & 0.00 & 0.00 & 0.00 & 2.65 & 2.56 & 2.61 \\ 1.40 & 1.41 & 1.38 & 1.38 & 1.41 & 1.40 & 1.40 & 1.38 & 1.40 & 1.39 & 1.41 & 1.41 & 1.41 & 1.40 & 1.38 & 1.39 & 2.80 & 0.00 & 2.67 & 2.66 & 0.00 & 2.77 & 2.77 & 2.67 & 2.77 & 2.70 & 0.00 & 0.00 & 0.00 & 2.79 & 2.69 & 2.74 \\ 1.48 & 1.49 & 1.46 & 1.46 & 1.49 & 1.47 & 1.47 & 1.46 & 1.47 & 1.46 & 1.49 & 1.49 & 1.49 & 1.48 & 1.46 & 1.47 & 2.93 & 0.00 & 2.78 & 2.78 & 0.00 & 2.90 & 2.90 & 2.79 & 2.90 & 2.81 & 0.00 & 0.00 & 0.00 & 2.91 & 2.80 & 2.86 \\ 1.56 & 1.57 & 1.54 & 1.54 & 1.57 & 1.55 & 1.55 & 1.54 & 1.55 & 1.54 & 1.57 & 1.57 & 1.57 & 1.56 & 1.54 & 1.55 & 3.05 & 0.00 & 2.89 & 2.89 & 0.00 & 3.02 & 3.02 & 2.90 & 3.02 & 2.92 & 0.00 & 0.00 & 0.00 & 3.03 & 2.91 & 2.97 \\ 1.64 & 1.66 & 1.62 & 1.62 & 1.66 & 1.64 & 1.64 & 1.62 & 1.64 & 1.63 & 1.66 & 1.66 & 1.66 & 1.64 & 1.63 & 1.63 & 3.17 & 0.00 & 3.00 & 3.00 & 0.00 & 3.13 & 3.13 & 3.01 & 3.13 & 3.03 & 0.00 & 0.00 & 0.00 & 3.15 & 3.02 & 3.08 \\ 1.74 & 1.75 & 1.72 & 1.72 & 1.75 & 1.73 & 1.73 & 1.72 & 1.73 & 1.72 & 1.75 & 1.75 & 1.75 & 1.73 & 1.72 & 1.72 & 3.30 & 0.00 & 3.12 & 3.12 & 0.00 & 3.25 & 3.25 & 3.12 & 3.25 & 3.14 & 0.00 & 0.00 & 0.00 & 3.27 & 3.13 & 3.20 \\ 1.84 & 1.86 & 1.82 & 1.82 & 1.86 & 1.83 & 1.83 & 1.82 & 1.83 & 1.82 & 1.86 & 1.86 & 1.86 & 1.84 & 1.82 & 1.83 & 3.44 & 0.00 & 3.24 & 3.24 & 0.00 & 3.37 & 3.37 & 3.25 & 3.37 & 3.27 & 0.00 & 0.00 & 0.00 & 3.40 & 3.26 & 3.32 \\ 1.97 & 1.98 & 1.95 & 1.95 & 1.98 & 1.96 & 1.96 & 1.95 & 1.96 & 1.95 & 1.98 & 1.98 & 1.98 & 1.96 & 1.95 & 1.95 & 3.60 & 0.00 & 3.39 & 3.39 & 0.00 & 3.52 & 3.52 & 3.39 & 3.52 & 3.41 & 0.00 & 0.00 & 0.00 & 3.55 & 3.40 & 3.47 \\ 2.12 & 2.14 & 2.10 & 2.10 & 2.14 & 2.12 & 2.12 & 2.10 & 2.12 & 2.11 & 2.14 & 2.14 & 2.14 & 2.12 & 2.10 & 2.11 & 3.80 & 0.00 & 3.56 & 3.56 & 0.00 & 3.69 & 3.69 & 3.56 & 3.69 & 3.58 & 0.00 & 0.00 & 0.00 & 3.74 & 3.57 & 3.64 \\ 2.33 & 2.35 & 2.31 & 2.31 & 2.35 & 2.32 & 2.32 & 2.31 & 2.32 & 2.31 & 2.35 & 2.35 & 2.35 & 2.33 & 2.31 & 2.32 & 4.05 & 0.00 & 3.79 & 3.79 & 0.00 & 3.92 & 3.92 & 3.79 & 3.92 & 3.80 & 0.00 & 0.00 & 0.00 & 3.97 & 3.80 & 3.87 \\ 2.63 & 2.65 & 2.61 & 2.61 & 2.65 & 2.62 & 2.62 & 2.61 & 2.62 & 2.61 & 2.65 & 2.65 & 2.65 & 2.63 & 2.61 & 2.62 & 4.41 & 0.00 & 4.11 & 4.11 & 0.00 & 4.23 & 4.23 & 4.11 & 4.23 & 4.12 & 0.00 & 0.00 & 0.00 & 4.30 & 4.11 & 4.18 \\ 3.09 & 3.11 & 3.07 & 3.07 & 3.11 & 3.08 & 3.08 & 3.07 & 3.08 & 3.07 & 3.11 & 3.11 & 3.11 & 3.09 & 3.07 & 3.08 & 4.95 & 0.00 & 4.57 & 4.57 & 0.00 & 4.68 & 4.68 & 4.57 & 4.68 & 4.58 & 0.00 & 0.00 & 0.00 & 4.77 & 4.58 & 4.65 \\ 3.85 & 3.86 & 3.82 & 3.82 & 3.86 & 3.83 & 3.83 & 3.82 & 3.83 & 3.82 & 3.86 & 3.86 & 3.86 & 3.84 & 3.82 & 3.83 & 5.81 & 0.00 & 5.31 & 5.31 & 0.00 & 5.39 & 5.39 & 5.31 & 5.39 & 5.31 & 0.00 & 0.00 & 0.00 & 5.51 & 5.31 & 5.38 \\ 5.17 & 5.17 & 5.13 & 5.13 & 5.17 & 5.13 & 5.13 & 5.13 & 5.13 & 5.13 & 5.17 & 5.17 & 5.17 & 5.14 & 5.13 & 5.13 & 7.28 & 0.00 & 6.50 & 6.50 & 0.00 & 6.56 & 6.56 & 6.50 & 6.56 & 6.51 & 0.00 & 0.00 & 0.00 & 6.69 & 6.51 & 6.57 \\ 7.47 & 7.43 & 7.39 & 7.39 & 7.43 & 7.39 & 7.39 & 7.39 & 7.39 & 7.39 & 7.43 & 7.43 & 7.43 & 7.41 & 7.39 & 7.40 & 9.77 & 0.00 & 8.42 & 8.42 & 0.00 & 8.44 & 8.44 & 8.42 & 8.44 & 8.42 & 0.00 & 0.00 & 0.00 & 8.54 & 8.42 & 8.47 \\ 11.25 & 11.04 & 11.02 & 11.02 & 11.04 & 11.02 & 11.02 & 11.02 & 11.02 & 11.02 & 11.04 & 11.04 & 11.04 & 11.03 & 11.02 & 11.02 & 13.82 & 0.00 & 11.23 & 11.23 & 0.00 & 11.23 & 11.23 & 11.23 & 11.23 & 11.23 & 0.00 & 0.00 & 0.00 & 11.26 & 11.23 & 11.25 \\ 16.35 & 15.75 & 15.75 & 15.75 & 15.75 & 15.75 & 15.75 & 15.75 & 15.75 & 15.75 & 15.75 & 15.75 & 15.75 & 15.75 & 15.75 & 15.75 & 20.23 & 0.00 & 15.16 & 15.16 & 0.00 & 15.16 & 15.16 & 15.16 & 15.16 & 15.16 & 0.00 & 0.00 & 0.00 & 15.16 & 15.16 & 15.16}$}}}
\end{table*}

\end{document}